\documentclass[aps,prd,superscriptaddress,showpacs,nofootinbib,amssymb,amsfonts,amssymb,amsmath,twocolumn,balancelastpage,floatfix]{revtex4-1}

\usepackage[utf8x]{inputenc}
\usepackage[dvipdf,dvips,dvipdfmx]{graphicx}
\usepackage{wrapfig}
\usepackage{bm}
\usepackage{braket}
\usepackage{color}
\usepackage{bbm}
\usepackage{xspace}
\usepackage{xfrac}
\usepackage{hyperref}
\usepackage[nameinlink]{cleveref}
\usepackage{xifthen}
\usepackage{xcolor}
\hypersetup{
	colorlinks,
	linkcolor={red!75!black},
	citecolor={blue!75!black},
	urlcolor={blue!75!black}
}

\newcommand{\gsim}{\gtrsim}

\newcommand{\p}{\partial}

\newcommand{\df}{\text{d}}
\newcommand{\Tr}{{\rm Tr}\,}
\newcommand{\pmat}[1]{\begin{pmatrix}#1\end{pmatrix}}

\graphicspath{{./figures/}}

\newcommand{\cblue}[1]{{\color{blue}#1}}

\newbox{\ORCIDicon}

\graphicspath{{./figures/}}
\sbox{\ORCIDicon}{\large\includegraphics[width=0.8em]{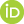}}
\allowdisplaybreaks[4]

\begin{document}

\title{
The $\theta$-vacuum from functional renormalisation
}

%%%%%%%%%%%%%%%%%%%%
\author{Yuepeng \surname{Guan}\,\href{https://orcid.org/0009-0007-8571-0931}{\usebox{\ORCIDicon}}\,}
\affiliation{Center for Theoretical Physics and College of Physics, Jilin University, Changchun 130012, China}
%%%%%%%%%%%%%%%%%%%%
\author{Jan M. \surname{Pawlowski}\,\href{https://orcid.org/0000-0003-0003-7180}{\usebox{\ORCIDicon}}\,}
\affiliation{Institut f{\" u}r Theoretische Physik, Universit{\" a}t Heidelberg, Philosophenweg 16, 69120 Heidelberg, Germany}
\affiliation{ExtreMe Matter Institute EMMI, GSI Helmholtzzentrum f{\" u}r Schwerionenforschung mbH, Planckstr.~1, 64291 Darmstadt, Germany}
%%%%%%%%%%%%%%%%%%%%
\author{Masatoshi \surname{Yamada}\,\href{https://orcid.org/0000-0002-1013-8631}{\usebox{\ORCIDicon}}\,}
\affiliation{Department of Physics and Astronomy, Kwansei Gakuin University, Sanda, Hyogo, 669-1330, Japan}
%%%%%%%%%%%%%%%%%%%%

\begin{abstract}
We study topological properties of a quantum mechanical system with $U(1)$-symmetry within the functional renormalisation group (fRG) approach. 
These properties include the vacuum energy structure and the topological susceptibility. 
Our approach works with a complexification of the flow equation, and specifically we embed the original symmetry into the complex plane, $U(1)\rightarrow \mathbb{C}$. 
We compute the effective potential of a given topological sector 
by restricting ourselves to field configurations with given generalised non-trivial Chern–Simons numbers. 
The full potential is directly constructed from these sector potentials. 
Our results compare well with the benchmark results obtained from solving the corresponding Schr\"odinger equation. 
\end{abstract}

\maketitle

%%%%%%%%%%%%%%%%%%%%%%%%%%%%%%%%%%%%%%%%%%%%%%%
\section{Introduction}

Topological properties in quantum field theories (QFTs) are determined by the global structure of the system, rather than its local details. 
These properties emerge in systems where fields or physical states exhibit non-trivial configurations constrained by topology, leading to a variety of rich physical phenomena. 
Notable examples include the Aharonov–Bohm effect~\cite{Aharonov:1959fk}, topological defects such as vortices and domain walls~\cite{Vilenkin:1984ib}, instanton-mediated tunneling effects~\cite{Belavin:1975fg, Coleman:1978ae}, and quantum anomalies~\cite{tHooft:1976rip}. 
These different phenomena highlight that non-perturbative approaches have to accommodate topological properties in a robust way.   

A non-perturbative functional approach is provided by the functional renormalisation group (fRG). 
It has been used for studying a large variety of non-perturbative phenomena, ranging from low-dimensional statistical systems over quantum chromodynamics (QCD) and beyond the Standard Model particle physics to quantum gravity, for a comprehensive recent review see \cite{Dupuis:2020fhh}. 
These applications also incorporate topological phenomena in quantum field theory and quantum mechanics; see e.g.~\cite{Reuter:1995tr,Reuter:1996be, Pawlowski:1996ch,VonGersdorff:2000kp, Kapoyannis:2000sp, Zappala:2001nv, Aoki:2002ozs,
Aoki:2002ax,Aoki:2002yt, Weyrauch:2006aj, Synatschke:2008pv, Jakubczyk:2016rvr, Krieg_2017, Bonanno:2022edf, Fukushima:2022zor, Bonanno:2025mon}. 
However, a comprehensive analysis of the respective dynamics is still lacking,  including an evaluation of the systematics in view of topological effects.    

This situation asks for a comprehensive analysis of the topological capacity of the fRG within a model theory, where the topological effects are well understood. Here we use $U(1)$-symmetric quantum mechanics, also called the quantum rotor model~\cite{Albandea:2024fui}. This model is a one-dimensional QFT that exhibits topological effects through the $\theta$-dependence of its partition function. It has been studied in a recent work \cite{Fukushima:2022zor}, where the fRG has been used in the local potential approximation, observing a lack of topological effects in this approximation. 

In the present work, we set up an fRG approach that accommodates topological effects. We show that the standard cutoff term for this model suppresses topological configurations for all cutoff scales, and only integrates out fluctuations. Accordingly, the information of the topological vacuum structure is already encoded in the UV boundary condition rather than being generated by the flow equation directly. In short, it is the appropriate combination of cutoff term and initial effective action that resolves the theory at hand. 

\begin{figure}
    \centering
\includegraphics[width=0.95\linewidth]{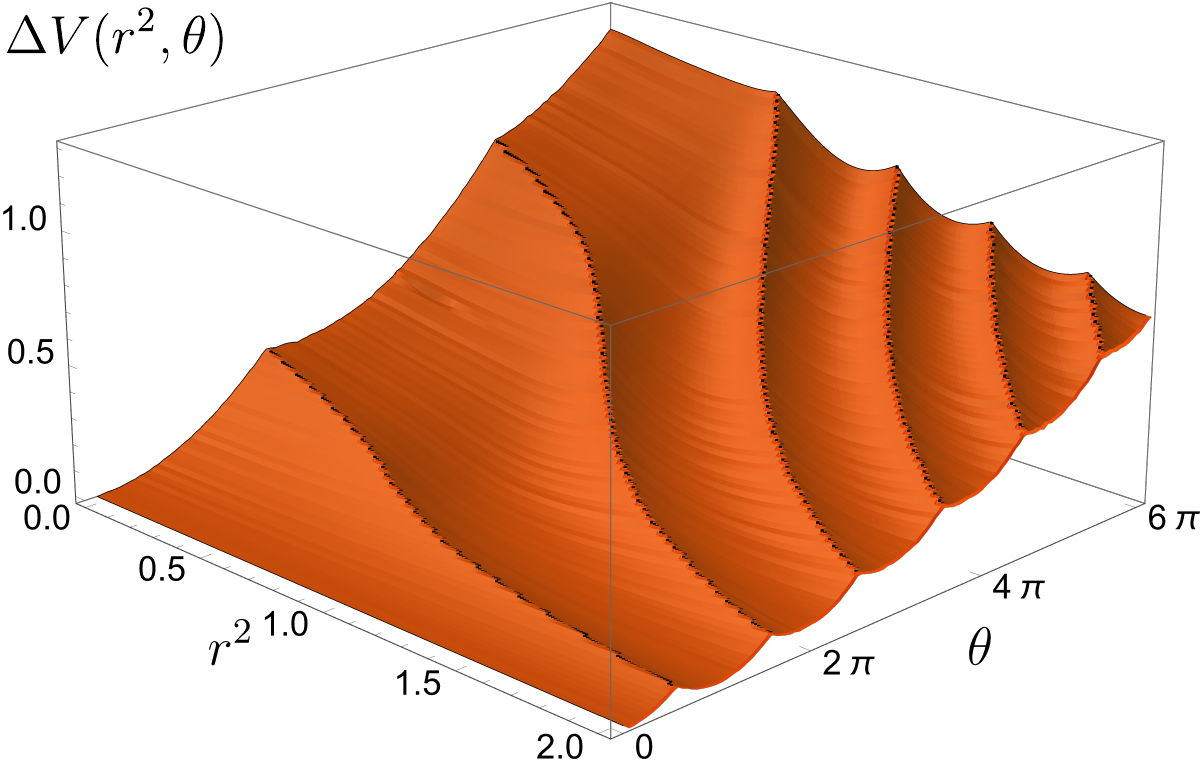}
    \caption{Effective potential $\Delta V(r^2,\theta)=V(r^2,\theta)-V(r^2,0)$, \labelcref{eq:DeltaV} of $U(1)$-symmetric quantum mechanics. It is obtained from the functional renormalisation group approach with Litim-type regulator, see \Cref{sec:NumericalResults}.}
    \label{fig:EffectivePotential}
\end{figure}

We confirm the practical feasibility of our approach by computing the vacuum structure of the model and, in particular, the vacuum energy and topological susceptibility. Our results, obtained in a local potential approximation, are in good agreement with the benchmark result by numerically solving the vacuum energy directly from the Schr\"odinger equation.
This is illustrated in \Cref{fig:EffectivePotential}, where we show as the effective potential of the semi-topological model obtained from the fRG approach set up here. This result illustrates the capacity of the fRG approach to accommodate the effects of topological configurations: The effective potential exhibits a non-analytic structure that originates in its semi-topological property, and hence the different topological sectors are clearly visible. 

This paper is organised as follows. 
In \Cref{sec:U1-QM}, we introduce the quantum rotor model and detail the theoretical deformation used to incorporate the $\theta$-dependence. We also provide benchmark results for the vacuum energy obtained by solving the Schr\"odinger equation. 
In \Cref{sec:fRG}, we introduce the fRG approach for the model. Specifically, we discuss, which degrees of freedom are suppressed by the cutoff term and derive the UV boundary condition for the effective action from the path integral. 
In \Cref{sec:NumericalResults}, we discuss our numerical results on the vacuum energy structure of this model as well as the topological susceptibility. 
In \Cref{sec:summary} we conclude, further derivations and supplementary computations are provided in the Appendices.

%%%%%%%%%%%%%%
\begin{figure*}
    \centering
    \includegraphics[width=0.47\linewidth]{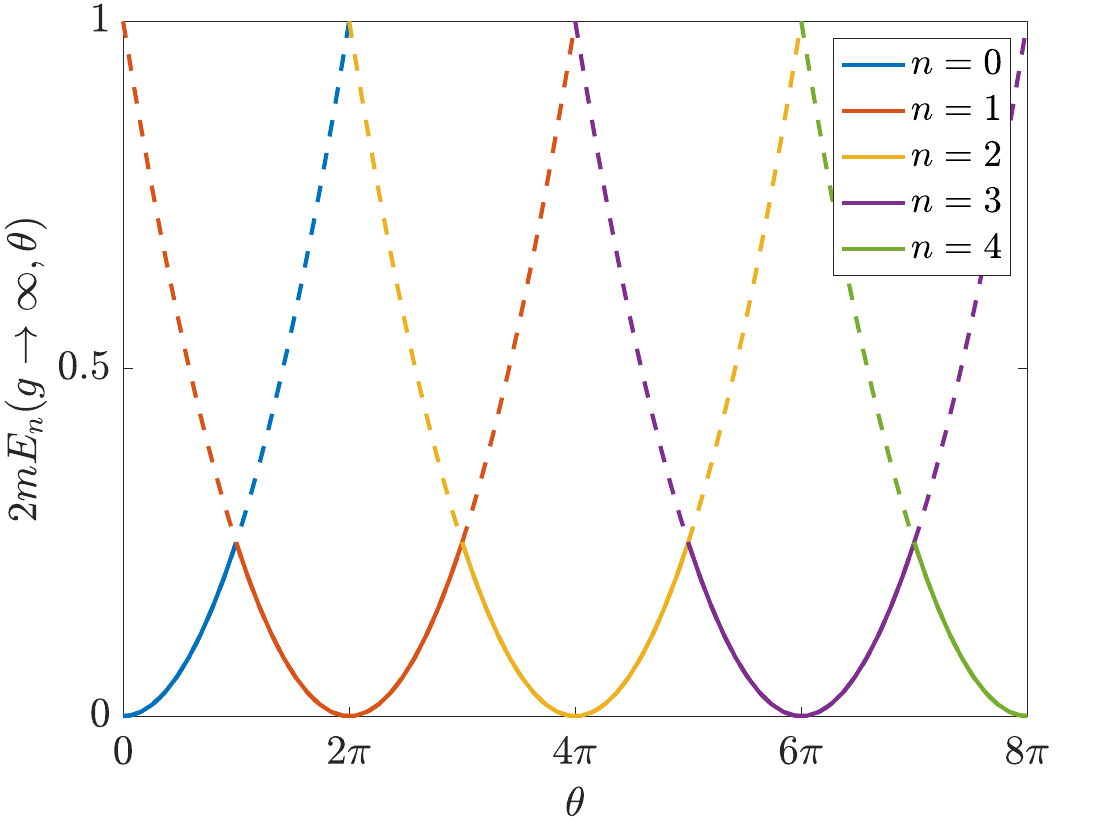}
    \includegraphics[width=0.45\linewidth]{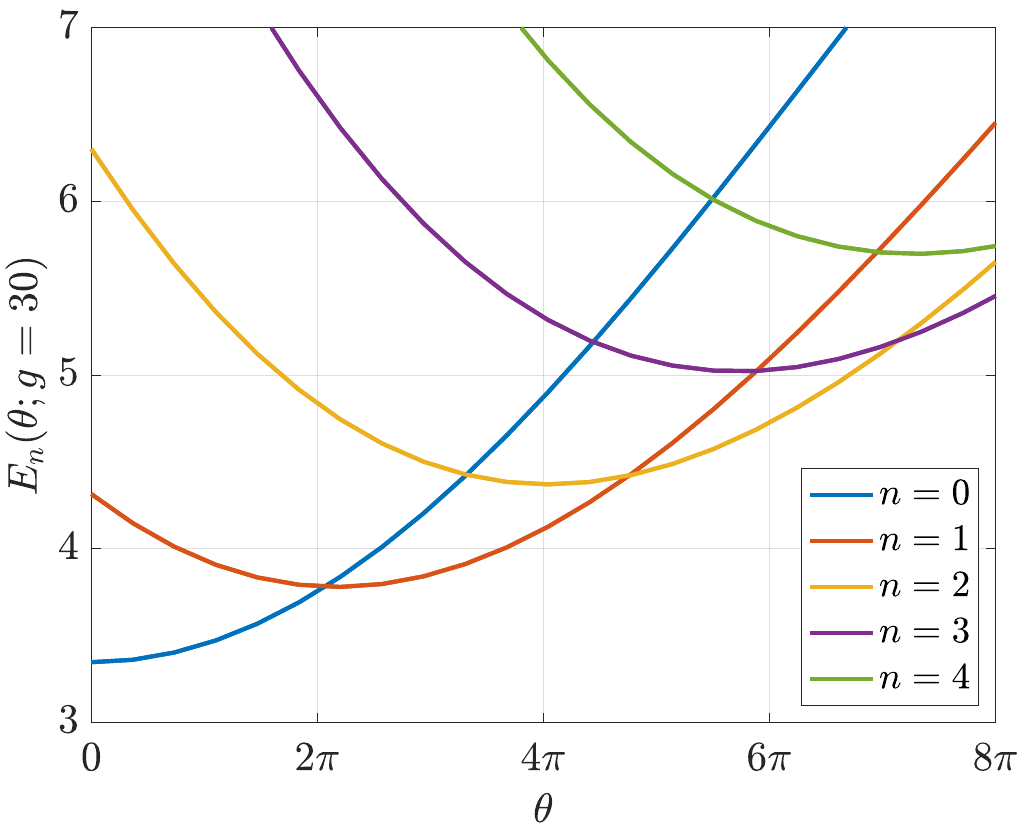}
    \caption{ $\theta$-parameter dependence of the energy levels for each $n$. The left-hand side panel exhibits the energy levels~\labelcref{eq:EgInfty} which corresponds to $g\to \infty$. The solid lines indicate the ground-state energy in the system.
    In the right-hand side panel, we depict the energy levels obtained by solving the Schr\"odinger equation~\labelcref{eq:RadialSchroedinger} for $g=30$. 
    }
    \label{fig:thetaVacuumExact}
\end{figure*}
%%%%%%%%%%%%%%

%%%%%%%%%%%%%%%%%%%%%%%%%%%%%%%%%%%%%%%%%%%%%%%%%%%%%%%%%
\section{\texorpdfstring{$U(1)$}{} symmetric quantum mechanics}
\label{sec:U1-QM}

In this Section, we introduce a $U(1)$-invariant quantum mechanical model with a topological $\theta$-term and briefly present its energy eigenvalue as a solution to the Schr\"odinger equation. Further details can be found in \Cref{app:QMwithTopTerm}. Its classical action is given by 
\begin{align} 
    S[\varphi] =  
    \int_\tau \, \left[ \frac{m}{2} \dot{\varphi}^* \dot{\varphi} - \frac{\theta}{4\pi} (\varphi^* \dot{\varphi} - \dot{\varphi}^* \varphi) + V(\varphi^*\varphi) \right]\,,
\label{eq:LagrangianUU}
\end{align}
where $\int_\tau = \int \mathrm{d}\tau$ denotes the integration over the Euclidean time $\tau$.
Here, $\varphi$ and $\varphi^*$ are quantum mechanical fields, and the potential is given by  
\begin{align}
    V(\varphi^*\varphi) = \frac{g}{4} (\varphi^* \varphi - 1)^2\,.
    \label{eq:originalInitialPotential}
\end{align}
This potential has a global minimum for fields with $\varphi^* \varphi = 1$.
The complex field $\varphi$ can be parametrised in terms of a polar basis, 
\begin{align}
\varphi=r\, e^{i\vartheta}\,,
\label{eq:uPolar}
\end{align} 
with real variables $r$ and $\vartheta$.  The potential \labelcref{eq:originalInitialPotential} only depends on the radial field $r$, and the action is written as
\begin{align}
S=\int_\tau\left[\frac{m}{2} ( \dot{r}^2 + r^2 \dot{\vartheta}^2 ) -i \frac{\theta}{2\pi} r^2 \dot{\vartheta} + \frac{g}{4} (r^2 - 1)^2 \right],
\end{align}
with $S=S[r,\vartheta]$. This concludes the setup of the model.

%%%%%%%%%%%%%%%%%%%%%%%%%%%%%%%%%
\subsection{Radial mean field limit: \texorpdfstring{$g\to\infty$}{}}
\label{sec:RadialMF}

We start our investigations with the study of the limit $g\to \infty$. In this limit, the $U(1)$-symmetric model reduces to the quantum rotor, also discussed in \Cref{app:QuantumRotor}: for $g\to\infty$ all radial fluctuations are suppressed and the path integral reduces to one over the angular degree of freedom $\vartheta(\tau)$: the potential turns into the delta function $\delta(\varphi\,\varphi^*-1)$ in the path integral, and we only have to consider field configurations with  
\begin{align}
    {\varphi}^* \varphi=r^2 = 1\,.
    \label{eq:uNorm1}
\end{align}
Then, the second term in \labelcref{eq:LagrangianUU}, the `$\theta$-term' becomes a topological term with 
\begin{align}
 \frac{\theta}{4\pi}\int_0^\beta \mathrm{d}\tau \, (\varphi^* \dot{\varphi} - \dot{\varphi}^* \varphi)
 &= \frac{\theta}{2\pi} \int_{\vartheta(0)}^{\vartheta(\beta)} \mathrm{d}\vartheta = i \theta \,\nu\,, 
 \label{eq:winding_number}
\end{align}
with the winding number $\nu\in\mathbbm{Z}$. 
Here, $\beta$ denotes the one-dimensional 'space-time' volume ($\tau \in [0, \beta]$), and the boundary condition for $\vartheta$ is given by 
\begin{align}
\vartheta(\beta) = \vartheta(0) + 2\pi \,\nu\,.
\label{eq:phi_boundary_condition}
\end{align}
In \labelcref{eq:phi_boundary_condition}, $\nu \in \mathbb{Z}$ is the winding number. We indicate the topological sector in the angular field with the subscript ${}_\nu$: $\vartheta_\nu$. This field is a sum of a sector-independent angular fluctuation $\vartheta=\vartheta_0$ and a global winding number part, 
\begin{align}
    \vartheta_\nu = \vartheta + 2 \pi\frac{\tau}{\beta} \, \nu\,.
    \label{eq:WindingSplitvartheta}
\end{align}
With \labelcref{eq:WindingSplitvartheta}, the field $\varphi$ in \labelcref{eq:uPolar} has the representation 
\begin{align}
\varphi(\tau)=\left[r\, e^{i\vartheta(\tau)}\right]\,e^{2 \pi\,i\,\frac{\tau}{\beta} \, \nu}\,. 
\label{eq:uPolarWinding}
\end{align} 
Although the topological term \labelcref{eq:winding_number} does not affect the classical equations of motion for $\varphi$ and $\varphi^*$ with $\varphi^*\varphi=1$, 
it contributes to the path integral through the summation over field configurations with different boundary conditions~\labelcref{eq:phi_boundary_condition}. 
Consequently, the energy level depends on $\theta$ such that
\begin{align}
    E_n(\theta;g\to\infty) = \frac{1}{2m} \left(n - \frac{\theta}{2\pi}\right)^2\,.
\label{eq:EgInfty}
\end{align}
Here, $n \in \mathbb{Z}$ is the quantum number labeling the energy spectrum, arising from the quantisation of angular fluctuations. For its derivation, see \Cref{app:QuantumRotor}. 
The ground-state energy in this system is periodic in $\theta$, as shown in the panel on the left side of \Cref{fig:thetaVacuumExact}.
This fact indicates the topological nature of this model. We also note for later use that the normalisation of the radial part to $r=1$ is a pure convention; the above derivation holds for any radial mean field value $r=\bar r$ instead of $r=1$. In the more general case, the energy levels \labelcref{eq:EgInfty} turn into 
\begin{align}
    E_n(\theta;g\to\infty) = \frac{1}{2m} \left(n - \frac{\theta\, \bar r^2}{2\pi}\right)^2\,. 
\label{eq:EgInftyGeneral}
\end{align}
This concludes our analysis of the radial mean field limit.

%%%%%%%%%%%%%%%%%%%%%%%%%%%%%%%%%
\subsection{General case with finite couplings \texorpdfstring{$g$}{}}
\label{app:Generalg}

The model for finite couplings $g$ is also discussed in \Cref{app:RelaxedQuantumRotor}. We solve the Schr\"odinger equation for $u=re^{i\vartheta}$ to obtain the energy levels. We use that the since the theory has $U(1)$ (or $O(2)$) symmetry. Therefore, we can write the wave function as a product of the radial and angle parts, i.e., 
\begin{align} 
\psi(r,\vartheta) = \sum_{l,n}\Upsilon_{l,n}(r)\Phi_n(\vartheta)\,.
\end{align} 
The angular variable $\vartheta$ does not enter the $\theta$-term and the potential. Hence, the solution of the  Schr\"odinger equation for the wave function $\Phi_n(\vartheta)$ is given by $\Phi_n(\vartheta)=e^{in\vartheta}$ up to an overall factor.
The Schr\"odinger equation for $\Upsilon_{l,n}(r)$ reads
\begin{align}\nonumber 
&    \Biggl[ -\frac{1}{2 m r} \frac{\partial}{\partial r}\left(r \frac{\partial}{\partial r} \right) + \frac{1}{2 m r^2} \left( n - \frac{\theta}{2 \pi} r^2 \right)^2 \\[1ex] 
& \hspace{1cm}+ \frac{g}{4} (r^2 - 1)^2 \Biggr] \Upsilon_{l,n}(r) = E_{l,n}(\theta;g) \Upsilon_{l,n}(r)\,,
    \label{eq:RadialSchroedinger}
\end{align}
see also \Cref{app:RelaxedQuantumRotor}. 
If solving this equation with the boundary condition $\Upsilon_{l,n}(r\to \infty) \to 0$, we obtain the energy eigenvalues $E_{l,n}(\theta;g)$. Here, $l$ stands for the quantum number corresponding to the radial excitation.
In this work, we consider the ground state $l=0$ and denote 
\begin{align} 
E_{n}(\theta;g):=E_{l=0,n}(\theta;g)\,.
\end{align} 
We illustrate some of the properties of the model with the parameters $g=30$ and $m=1$. For these parameters the dependence of the energy levels on the $\theta$-parameter is shown in the right-hand side panel of \Cref{fig:thetaVacuumExact}.
We see that the periodicity of the energy eigenvalues with respect to changes in $\theta$ is lost.
Nonetheless, we still observe the level crossing of the ground-state energy.
Note that for $g=0$, the energy eigenvalues are given by
\begin{align}
E_n(\theta;g=0) = \frac{|\theta|}{\pi m} \left( n + \frac{1}{2}\right)\,.
\label{eq:zeropoint energy}
\end{align}
We also note that for a large finite value of $g$, the Schr\"odinger equation~\labelcref{eq:RadialSchroedinger} involves the zero-point energy $E^{(0)}$, which is proportional to $\sqrt{g}$, 
\begin{align}
E^{(0)}(g) =\sqrt{\frac{g}{2}}\,.
\end{align}

In the following Sections, we use the functional renormalisation group (fRG) to compute the energy eigenvalues of the system \labelcref{eq:LagrangianUU}. This serves as a benchmark test for the capacity of the fRG to resolve topological properties of quantum field theories, for a respective benchmark computation in the anharmonic oscillator see \cite{Bonanno:2025mon}.

%%%%%%%%%%%%%%%%%%%%%%%%%%%%%%%%%%
\section{Functional renormalisation group approach to \texorpdfstring{$U(1)$}{}-symmetric quantum mechanics}
\label{sec:fRG}

In the present work we use the functional renormalisation group approach for computing the non-perturbative effective potential of $U(1)$-symmetric quantum mechanics. Specifically, we use the flow equation for the quantum effective action, the Wetterich equation \cite{Wetterich:1992yh}, see also \cite{Ellwanger:1993mw, Morris:1993qb}. For a recent comprehensive review see~\cite{Dupuis:2020fhh}. 

In the following, we will consider the Cartesian field $\phi=\varphi$ as well as polar coordinates \labelcref{eq:uPolar} with $\phi=r,\vartheta$. 
The respective flow equations for the effective actions $\Gamma_\phi$ are derived from infrared regularised generating functionals with infrared cutoff terms and current terms for the respective fields. We write 
\begin{align} 
Z_\phi[J_\phi] = \int {\cal D}\hat\varphi\, e^{-S[\hat\varphi]+\Delta S_\phi[\hat\phi]+ \int_\tau J^a_\phi \hat\phi^a}\,, 
\label{eq:Zphi}
\end{align} 
where the cutoff term $\Delta S_\phi[\hat\phi]$ is quadratic in the field that is coupled to the source, 
\begin{align}
    \Delta S_k[\phi] = \frac12 \int_p \phi_a(-p)\, R^{ab}_k(p)\,\phi_b(p)  \,.
\label{eq:DeltaS}
\end{align}
Hence, the dispersion of the field $\phi$ receives an additional contribution $R_k(p)$, that can be understood as a momentum-dependent mass term. For this more general framework see \cite{Pawlowski:2005xe, Ihssen:2024ihp}. 

In the present work we mostly consider the Cartesian field $\hat\phi=\hat\varphi$, specifically within the numerical applications. However, for the discussion of potential regularisation of the topological degrees of freedom we also consider the polar coordinates $\hat\phi[\hat\varphi]=(\hat r,\hat\vartheta)$.

%%%%%%%%%%%%%%%%%%%%%%%%%%%%%%%%%%
\subsection{Flow equation}
\label{sec:Flow}

For $U(1)$-symmetric quantum mechanics, the flow equation for the effective action, the Legendre transform of $\log Z_\phi[J_\phi]$,  reads  
\begin{align}
    \partial_t \Gamma_k[\phi] = \frac{1}{2} \int_{q,p}  \left[  \frac{1}{\Gamma_k^{(2)}[\phi] + R_k}\right]_{ba}\hspace{-.2cm}(-p,p)\,\partial_t R^{ab}_k(p)\,. 
\label{eq:wettericheq}
\end{align}
In \labelcref{eq:wettericheq} we have dropped the subscript ${}_\phi$, but it is understood that $\Gamma=\Gamma_\phi$ differs for different fields coupled to the current. The RG-time $t = \log (k/\Lambda)$ is the logarithmic infrared cutoff scale, measured in a reference scale $\Lambda$. In the following, we choose $\Lambda$ to be the initial cutoff scale, from which the flow is integrated down to $k=0$. In \labelcref{eq:wettericheq}, $\Gamma_k^{(2)}$ is the full two-point function, 
\begin{align}
  \Gamma_{k,ab}^{(2)}(p,q)= \frac{\delta^2\Gamma_k }{\delta \phi^a(p)\,\delta\phi^b(q)} \,. 
  \label{eq:Gamma2}
\end{align}
In this setup the complete effective action $\tilde \Gamma_k[\phi]$ is given by $\Gamma_k[\phi]+\Delta S_k[\phi]$ and the first part satisfies the flow equation \labelcref{eq:wettericheq}.  
The full quantum effective action is obtained by integrating the flow of the effective action from an initial scale $k=\Lambda$ to $k=0$, to wit, 
\begin{align} 
\Gamma[\phi] = \Gamma_\Lambda[\phi] + \int_\Lambda^0 \frac{\df k}{k} \partial_t \Gamma_k[\phi]\,.
\label{eq:IntegratedFlow}
\end{align} 
The approach is completed with an initial condition $\Gamma_\Lambda$ at an asymptotically large scale, where the theory tends towards the ultraviolet (bare) action, 
\begin{align} 
\Gamma_{\Lambda\to \infty}[\phi]  = S\bigl[\varphi[\phi]\bigr] +\Delta \Gamma_\textrm{UV}[\phi]\,. 
\end{align} 
In a general theory in $d$ dimensions the additional term $\Delta \Gamma_\textrm{UV}[\phi]$ comprises UV-relevant terms other than that already present in the classical action. Such terms may be generated by the presence of the cutoff term itself. A well-known example is the gluon mass term in QCD, that is generated by the breaking of gauge symmetry in the presence of the cutoff.

%%%%%%%%%%%%%%%%%%%%%%%%%%%%%%%%%%
\subsection{Topology, flows and initial conditions} 
\label{sec:Initial+Top}

An additional intricacy for the initial condition emerges in the presence of topological configurations. A showcase relevant for our present study are anomalous $U(1)$-violating terms in QCD (the 't Hooft determinant) that are generated by the fermionic path integral measure. This case has been studied in \cite{Pawlowski:1996ch} within the fRG. It has been argued there that for chirally-symmetric cutoffs no anomalous contribution to the effective action is generated in the chiral limit if this term is not sourced in the initial effective action. Indeed, the situation is not completely resolved and one of the motivations of the present study is to shed some light into these intricacies. In any case, the study in \cite{Pawlowski:1996ch} suggests separating topological configurations from fluctuations that are the same in each sector. Then, the fluctuations are infrared regularised while the instantons are summed over without any restriction.

%%%%%%%%%%%%%%%%%%%%%%%%%%%%
\subsubsection{Fluctuation field regulators}
\label{sec:FlucFieldRegs}

It is suggestive that the mechanisms mentioned above are also present in the quantum mechanical system under investigation here: For illustration, we first consider a polar decomposition 
\begin{align} 
\hat\varphi_\nu= \hat r \,e^{i \hat \vartheta_\nu}\,,
\qquad
\varphi_{\bar \nu} = 
\langle \hat\varphi_\nu\rangle =r\,e^{i \vartheta_{\bar \nu}}\,. 
\label{eq:Flucrvartheta}
\end{align}
The angular mean field $\vartheta$ may also carry a winding number $\bar\nu$, see \labelcref{eq:WindingSplitvartheta}. Consequently, such a basis choice facilitates the separation and analysis of the topological configurations. We use polar cutoff terms, 
\begin{align} 
\Delta S_{\textrm{pol}}[r,\vartheta_\nu] =\Delta S_{r}[r]+  \Delta S_{\vartheta}[\vartheta_\nu]\,, 
\label{eq:PolarCutoff}
\end{align}
with 
\begin{align} \nonumber 
\Delta S_{r}[r] = &\,\frac12 \int_\tau r(\tau) R_r(\partial^2_\tau)  r(\tau)\,, \\[1ex]
\Delta S_{\vartheta}[\vartheta_\nu] = &\,\frac12 \int_\tau  \vartheta_\nu(\tau) 
R_\vartheta(\partial^2_\tau)  \vartheta_\nu(\tau)\,.  
\label{eq:PolarRegs}
\end{align}
For the time being, we have used the full angular mode in the cutoff term. With the decomposition \labelcref{eq:WindingSplitvartheta} the angular part reads 
\begin{align} 
\Delta S_{\vartheta}[\vartheta_\nu] = \Delta S_{\vartheta}[\vartheta] + \frac{2 \pi^2}{3} \nu \beta\,k \,.  
\label{eq:PolarRegSplit}
\end{align} 
The second term on the right hand side of \labelcref{eq:PolarRegSplit} comes from the integration of the winding mode squared, $4 \pi \nu^2 \tau^2 /\beta^2$ with $R_\vartheta(0)=k$. In the following calculation we only consider the limit $\beta\to \infty$ and the cutoff term \labelcref{eq:PolarRegSplit} suppresses all winding $\nu\neq 0$ for all $k>0$. Accordingly, the respective flow will not reproduce the topological sum. We note that this result also holds for cutoffs in $\varphi$, but is most transparent in the polar decomposition.

%%%%%%%%%%%%%%%%%%%%%%%%%%%%
\subsubsection{Initial effective action}
\label{sec:InitialEffAct}

The discussion in \Cref{sec:FlucFieldRegs} indicates that the initial effective action may or may not contain topological terms subject to the chosen cutoff. Moreover, specific cutoff choices may also introduce what is known as 'topological freezing' on the lattice. For the investigation of these intricacies we consider the integro-differential form of the effective action of the Cartesian fields $\varphi$, 
\begin{align}\nonumber 
e^{-\Gamma_k[\varphi]} = &\,\lim_{\beta\to \infty}\sum_\nu\int \mathcal D\hat\varphi\,e^{-S\left[\hat\varphi\,e^{  \frac{2\pi i \nu}{\beta} \tau}\right] +\Delta S_{k}[\hat\varphi-\varphi]} 
\\[1ex] 
& \hspace{2.5cm}\times e^{ \int_\tau \left(\hat \varphi -\varphi\right) \frac{\delta\Gamma[\varphi]}{\delta \varphi } }\,. 
\label{eq:GkPI}
\end{align}
Assume now that $\varphi$ is a fluctuation mean field in the trivial topology sector. Then, even though $[\mathcal D\hat\varphi]$ integrated over all fields, it is reduced to the measure in the trivial topology sector due to the regulator term: for $\hat\varphi$ with non-trivial windings the regulator term diverges in the limit $\beta\to \infty$. 

In the limit $k\to \infty$, the integration of the fluctuation field is suppressed and it converges to the mean field, $\hat\varphi\to \varphi$. This leads us to  
\begin{align}
\lim_{k\to \infty} e^{-\Gamma_k[\varphi]} 
\propto  &\sum_{\nu} e^{-S\left[\varphi\,e^{  \frac{2\pi i\nu}{\beta} \tau}\right] }\,. 
\label{eq:GkPIInfty}
\end{align}
The classical action is given by   
\begin{align}
 S\left[\varphi\,e^{  \frac{2\pi i \nu}{\beta} \tau}\right] = S(\varphi) +\left[ \frac{m}{2}\frac{(2\pi \nu)^2}{\beta} - i\nu\theta \right]  r^2\,, 
 \label{eq:TopSplitS}
\end{align}
where $r$ and $\vartheta$ are defined as the radial and angular part of the mean field $\varphi$. We emphasise that they are not the mean fields of $\hat r$ and $\hat\vartheta$. 

\begin{figure}
    \centering
\includegraphics[width=1\linewidth]{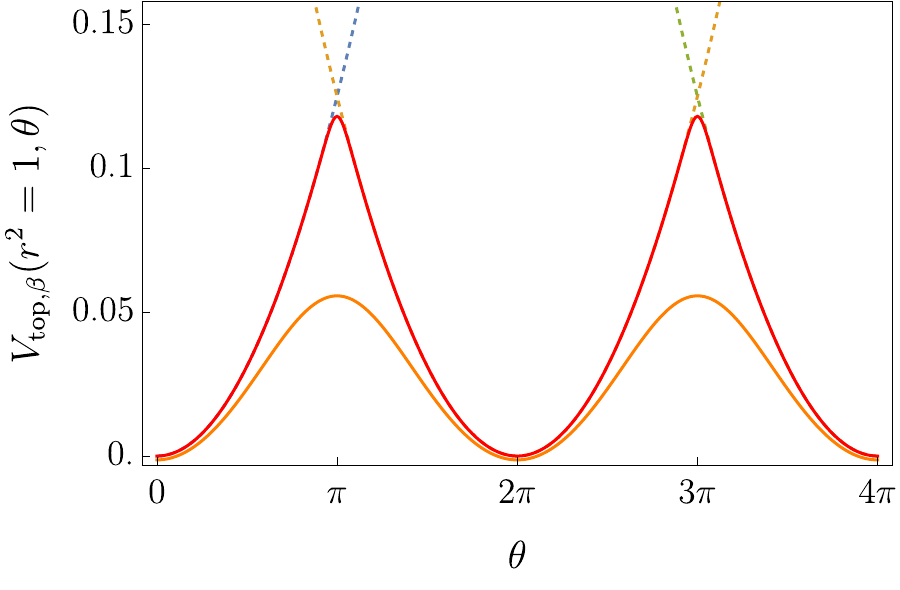}
    \caption{The additional potential energy $V_\textrm{top}$ with respect to the $\theta$-parameter at fixed $r^2 = 1$, which corresponds to the ``vacuum energy'' at finite ``temperature'' $1/\beta$. The dashed lines denote the potential energy with different energy levels $n$ with respect to $\theta$. The orange and the red lines denote the potentials with $\beta = 10$ and $\beta = 10^2$, respectively.}
    \label{fig:SummedVtopAtUVwrttheta}
\end{figure}
In \labelcref{eq:TopSplitS} we have assumed a constant $r$ to pull out the radial field in
\begin{align}
\frac{2 \pi \nu}{\beta} r^2 \int_\tau   \partial_\tau  \hat\vartheta = 0\,.  
\end{align}
Moreover, the $\tau$-integration in the winding number part is readily performed, leading to 
\begin{align}
    \int\limits_\tau \left[ \frac{m}{2}\frac{(2\pi \nu)^2}{\beta^2} - i\frac{\nu}{\beta}\theta \right] r^2= \left[ \frac{m}{2}\frac{(2\pi \nu)^2}{\beta} - i\nu\theta \right]  r^2\,.
\end{align}
The sum over the topological sectors can be performed analytically and we arrive at $k=\Lambda\to\infty$, 
\begin{subequations} 
\label{eq:ZFull}
\begin{align}
\Gamma_\Lambda[\varphi]  \simeq S[\varphi]+\beta\,  V_{\textrm{top},\beta}(r^2;\theta)\,. 
\label{eq:ZafterPoisson}
\end{align}
The 'topological' potential reads 
\begin{align}
    V_{\textrm{top},\beta}(r^2;\theta) = - \frac{1}{\beta} \log \left[ \sqrt{\frac{2 \pi m r^2}{\beta}} \vartheta_3 \left( -\frac{\theta}{2} r^2 , e^{\frac{2m \pi^2}{\beta} r^2}\right)  \right]\,,
    \label{eq:summed potential}
\end{align}
\end{subequations}
where $ \vartheta_3(z,q) $ denotes the Jacobi theta function of the third kind.
\begin{figure}
    \centering
\includegraphics[width=1\linewidth]{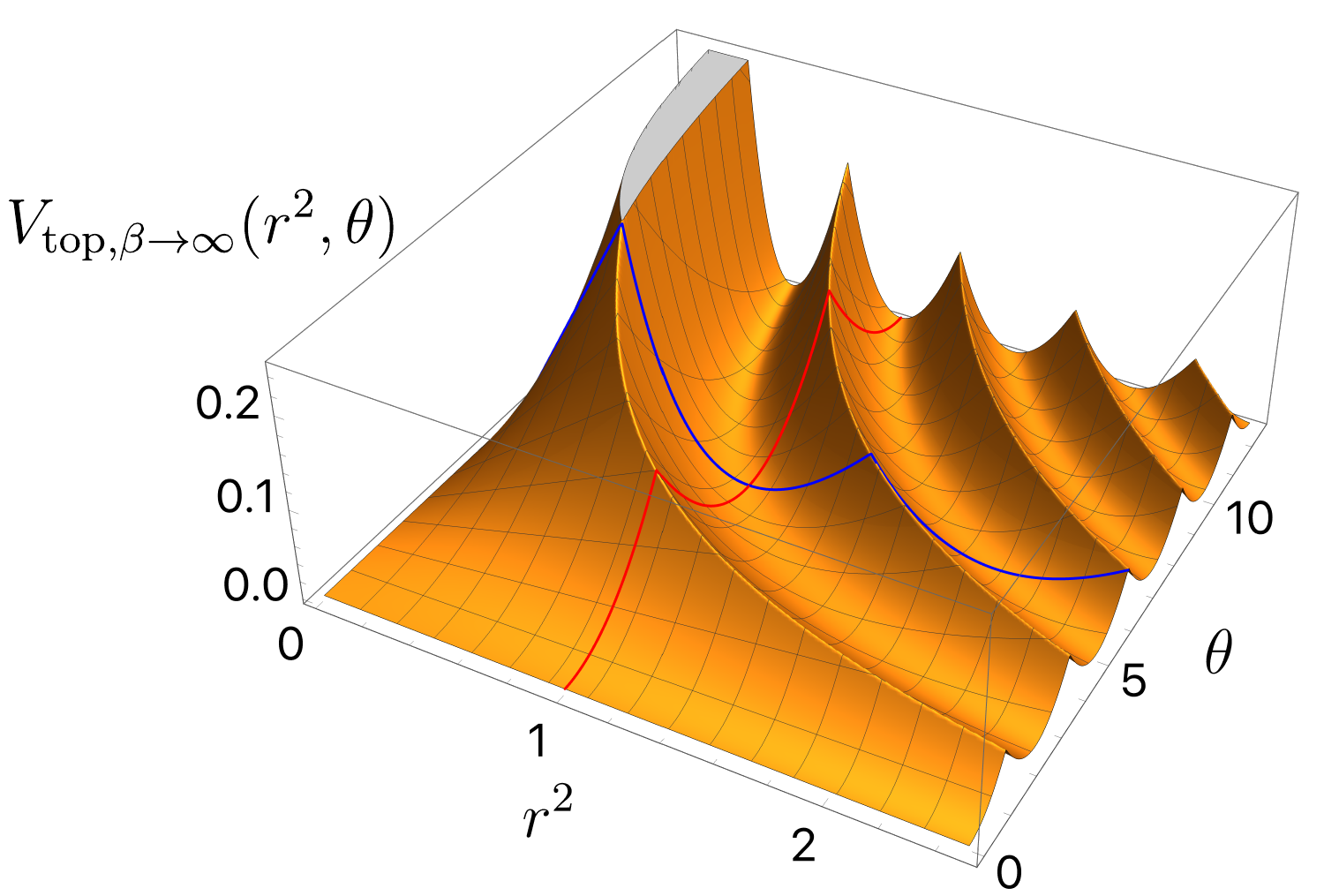}
    \caption{The additional potential $V_\textrm{top}(r^2,\theta)$ realised from the summation of all winding configurations. The red line denotes the potential evaluated at $r^2 = 1$, which corresponds to the ground-state energy (the left-hand side panel of \Cref{fig:thetaVacuumExact}), and the blue line denotes the potential in terms of the field variable with fixed $\theta = 2 \pi$.}
    \label{fig:SummedVtopAtUV}
\end{figure}
We show the potential \labelcref{eq:summed potential} as a function of $\theta$ for $r^2=1$ for different inverse temperatures $\beta$ \cblue{in \Cref{fig:SummedVtopAtUVwrttheta}}. In particular, in the limit $\beta\to \infty$ we obtain 
\begin{align}\nonumber 
  V_\textrm{top}(r^2;\theta)  =&   \lim_{\beta\to \infty} V_{\textrm{top},\beta}(r^2;\theta)\\[1ex]
   =& \frac{1}{2m} \left( \frac{\theta}{2 \pi} r^2  -\frac{n(\theta,r)}{r}  \right)^2\,,
  \label{eq:VtopAtUV}
\end{align}
with
\begin{align}
    n(\theta,r) = \operatorname{Floor} \left[ \frac{\theta}{2 \pi} r^2 + \frac{1}{2} \right]\,.
    \label{eq:energyLevelwrtr2theta}
\end{align}
The potential~\labelcref{eq:VtopAtUV} in the  $r^2$-$\theta$ plane is shown in \Cref{fig:SummedVtopAtUV}, for the slice $V_{\textrm{top},\beta}(r^2,\theta = 2 \pi)$ see  \Cref{fig:SummedVtopAtUVwrtr2}. 
\Cref{eq:VtopAtUV} entails that this potential has several non-analytical cusps that separate the $r^2$-$\theta$-plane into different branches.
These cusps originate from the transition between different ``energy levels'' \labelcref{eq:energyLevelwrtr2theta} when taking the limit $\beta \rightarrow \infty$. 
There, they can be read off as 
\begin{align}
    \theta \, r^2 = (2 n - 1) \pi, \quad n \in \mathbb{Z}\,.
    \label{eq:quantumnumber}
\end{align}
For a given $\theta$-parameter, the fRG flow for the effective potential starts with the initial potential combined with the potential \labelcref{eq:VtopAtUV} and the original one \labelcref{eq:originalInitialPotential}, with non-analyticities. 

\begin{figure}
    \centering
    \includegraphics[width=1\linewidth]{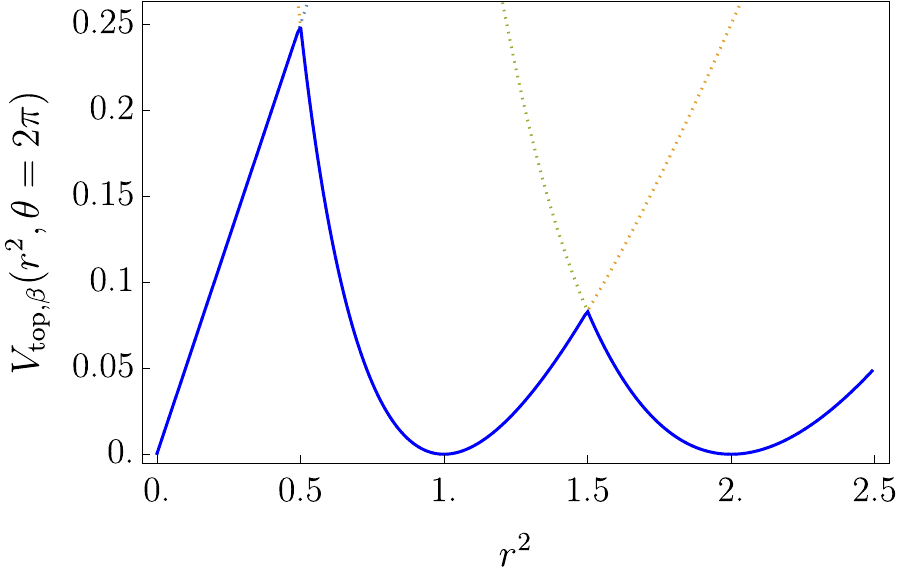}
    \caption{The $r$-dependence of the 'topological' potential energy $V_\textrm{top}(r^2,\theta$ fixed $\theta = 2 \pi$ and the limit $\beta \rightarrow \infty$ (solid blue line), which corresponds to the blue line in \Cref{fig:SummedVtopAtUV}. The dashed lines denote the potential energy \labelcref{eq:VtopAtUV} with different energy levels $n$, evaluated at $\theta = 2 \pi$.}
    \label{fig:SummedVtopAtUVwrtr2}
\end{figure}
%

%%%%%%%%%%%%%%%%%%%%%%%%%%%%%%%%%%%%%%
\subsubsection{Functional flows with topology}
\label{sec:FlowWithTopology}

In summary, the fRG approach with cutoff terms for the full fields,  including the winding part,  experiences a ``topology freeze'' in the limit $\beta\to \infty$: its naïve implementation does not incorporate the different topological sectors as shown in \cite{Fukushima:2022zor}. In turn, for a fixed finite $\beta$, the potential is periodic and the effective potential at $k=0$ is flat due to the convexity constraint. 

These intricacies are overcome by only regularising the fluctuation fields, leaving the windings unconstrained. The flow equation then follows readily from the integro-differential form of the effective action in \labelcref{eq:GkPI}. Applying a RG-time derivative to \labelcref{eq:GkPI} leads us to the standard flow equation 
\begin{align} 
\partial_t \Gamma_k[\varphi] = \frac12  \textrm{Tr} \left[ \frac{1}{ \Gamma_k^{(2)}[\varphi] + R_k } \partial_t R_k\right] \,.  
\end{align} 
The sum over the topological sector informs the initial condition \labelcref{eq:ZFull}, and we only consider the limit $\beta\to \infty$ with the potential $V_\textrm{top}(r^2;\theta)$ given by \labelcref{eq:VtopAtUV}.

%%%%%%%%%%%%%%%%%%%%%%%%%%%%%%
\subsection{Flow equation of the effective potential}
\label{sec:fRG_setup}

We use the local potential approximation (LPA), where only quantum corrections for constant fields are considered. The effective action reads  
\begin{align}
    \Gamma_k[\varphi] 
    = \int \frac{\df p}{2\pi} \left[ \frac{m}{2} \varphi^*(-p) p^2 \varphi(p)\right] + \int_\tau V_k(\varphi^*\varphi) \,, 
    \label{eq:GammakWithTopResum}
\end{align}
and the derivation of the flow equation for the effective potential is deferred to \Cref{sec:DerivationFlowApp}. We arrive at 
\begin{align}
\partial_t V_k
    &= \int \frac{\df p}{2\pi} \frac{\frac{m}{2}\partial_t {R}_k(p) \left[ \frac{m}{2}P_k(p) + V^{\prime} + r^2 V^{\prime\prime} \right]}{\left[ \frac{m}{2}P_k(p) + V^{\prime} + r^2 V^{\prime\prime} \right]^2 - \left[ r^2 V^{\prime\prime} \right]^2}\,,
    \label{eq:FlowVkMomentum}
\end{align}
with 
\begin{align}
P_k(p) = p^2 + R_k(p).
\end{align}
The results in the present work are obtained with the flat or Litim regulator, which has been shown to be the optimal regulator in LPA, see \cite{Litim:2001up}, 
\begin{align}
     R_k(p) = m \,(k^2-p^2)\Theta(k^2-p^2)\,,
     \label{eq:LitimReg}
\end{align}
where $\Theta(x)$ is the step function, 
\begin{align}
    \Theta(x) = \begin{cases}
        1, &  x \geq 0\,, \\[1ex]
        0, &  x < 0\,.
    \end{cases}
    \label{eq:Step}
\end{align}
Using \labelcref{eq:LitimReg}, we arrive at the following analytic flow equation for the effective potential, see \labelcref{eq:FlowVApp}.  
\begin{align}
    \partial_t V_k = \frac{1}{\pi} \frac{2\,m \,k^3\,\left( m \,k^2 + 2V_k^\prime + 2r^2V_k^{\prime\prime}\right)}{(m \,k^2 + 2V_k^\prime)^2 + 4r^2V_k^{\prime\prime}(m\, k^2 + 2V_k^\prime) }\,.
\label{LitimflowLPA}
\end{align}
In \Cref{app:ResultsCS} we also provide reference results obtained with the Callan-Symanzik (CS) type regulator $R_k(p)=k^2$: Firstly, from the computational point of view the CS-regulator is far away from an optimal one in LPA and hence provides a systematic error estimate for the fRG computations. Secondly, it keeps the analytic structure in frequency space intact, which is deformed by the non-analytic Litim regulator. The latter property makes it amiable towards direct real-time computations in quantum mechanics and beyond, see \cite{Fehre:2021eob, Braun:2022mgx}. 

In \Cref{sec:BreakdownLC}, we present two cases without topological resummation. First, we summarize the results from \cite{Fukushima:2022zor}. As an alternative setup, we consider the effective action with a complex frequency. In both cases, the corresponding flow equations fail to capture the level crossing of the ground-state energy.

%%%%%%%%%%%%%%%%%%%%%%%%%%%%%%%%%%%%%%%%%%%%%
\section{Numerical Results}
\label{sec:NumericalResults}

In this Section, we investigate the energy levels by solving the flow equation \labelcref{LitimflowLPA} and the Schr\"odinger equation \labelcref{eq:RadialSchroedinger} numerically and compare them. Moreover, all computations are done in the limit 
\begin{align} 
\beta\to \infty\,.
\end{align} 
In addition to the energy levels, we evaluate the topological susceptibility to examine whether or not the system tends to exhibit topological quantum mechanics for $g\to \infty$.

%%%%%%%%%%%%%%%%%%%%%%%%%%%%%%%%%%
\subsection{Energy levels}
\label{sec:EnergyLevelsWithout}

We first initiate the system in the ultraviolet with a fixed topological number (or quantum number of angular excitations) and ignore the cusps in the initial condition \labelcref{eq:ZFull} with \labelcref{eq:VtopAtUV}. This provides us with a potential $V_n(r^2,\theta)$, which is obtained by solving the flow equation~\labelcref{LitimflowLPA}  with an initial potential with a fixed $n$, 
\begin{align}
V_{n,k=\Lambda}(r^2)=  \frac{1}{2 m r^2} \left( n - \frac{\theta}{2 \pi} r^2 \right)^2 + \frac{g}{4} (r^2 - 1)^2\,,
\label{eq:initialConditionWithFixedn}
\end{align}
where the $\theta$-term is obtained from~\labelcref{eq:VtopAtUV} with   replacement $n(\theta,r^2)\to n$ in \labelcref{eq:energyLevelwrtr2theta} with \labelcref{eq:quantumnumber}. 

In \labelcref{eq:initialConditionWithFixedn} we have dropped an additive constant linear in $k$. 
This term takes into account that the overall normalisation of the path integral is $k$-dependent and has to be chosen such that at $k=0$ the states are normalised to unity: $\langle 1\rangle =1$. 
Then, the effective potential will not only allow us to compute energy difference but also the vacuum energy. 
The constant term does not feed back into the flow and we simply remove the respective contributions in the flow, 
\begin{align}
\p_t  V_{n,k} \to  \p_t V_{n,k} - \p_t V_{k}[V^\prime=0= V^{\prime\prime}]\,. 
\label{eq:ToDeltaV}
\end{align}
Evidently, the subtraction in \labelcref{eq:ToDeltaV} is independent of the potential $V_k$, and hence on the energy level. With a slight abuse of notation we still call the subtracted  potential $V_k$.  

The subtraction in~\labelcref{eq:ToDeltaV} removes the leading cutoff dependence of the field-independent part of the potentials $V_{n,k}$. The regulator also induces further ones which lower powers of $k$. Note however that they may also feature a $\theta$- and $n$-dependence which is absent in the leading order term. This will be discussed elsewhere. In the current work we circumvent this issue as follows: we normalise the fRG results for the energy levels $E_n(\theta;g)$  
to that from the Schr\"odinger equations at the minimum $\theta_{\textrm{min},n}$.

In summary, we obtain the energy levels from the subtracted potential $V_k$,  
\begin{align}\nonumber 
    E^\textrm{fRG}_n(\theta;g) =&\, V_{n,k=0}(\langle r\rangle^2,\theta)-V_{n,k=0}(\langle r\rangle^2,\theta_{\textrm{min},n})\\[1ex]
    &+E^\textrm{Ham}_n(\theta_{\textrm{min},n};g)\,. 
    \label{eq:Energy-fRG}
\end{align}
Here, $\langle r\rangle$ denotes the minimum of the effective potential and $E^\textrm{fRG/Ham}$ denotes the energy levels computed with the fRG and the Schr\"odinger equation respectively.  
Finally, we note that in quantum mechanical systems (one-dimensional quantum field theories), we always observe $\langle r\rangle=0$ thanks to the Hohenberg-Mermin-Wagner theorem~\cite{Mermin:1966fe, Hohenberg:1967zz}, see also Berezinskii and Coleman \cite{Berezinsky:1970fr, Coleman:1973ci}. \cref{eq:Energy-fRG}

\begin{figure}
    \centering
    \includegraphics[width=\linewidth]{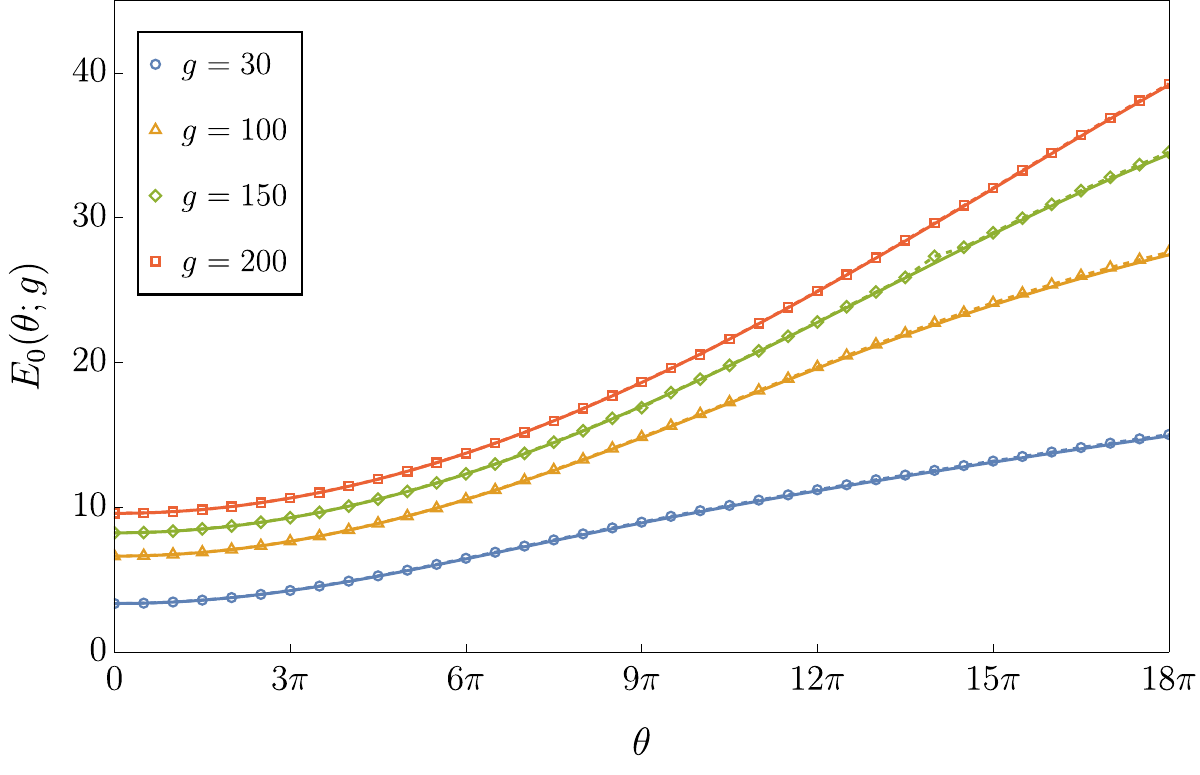}
    \caption{Energy levels~\labelcref{eq:Energy-fRG} for $n=0$ and different couplings $g$ in terms of the $\theta$-parameter from the flow equations with the Litim-type regulator. 
    The solid lines show the energy levels obtained from the Schr\"odinger equation~\labelcref{eq:RadialSchroedinger}.}
    \label{fig:n0energylevel}
\end{figure}
First, we investigate the energy eigenvalue for $n=0$ with several values of $g$.
In \Cref{fig:n0energylevel}, we show the $\theta$-dependence of the energy levels ($n=0$) for several values of $g$, which are obtained from the flow equation with the Litim regulator~\labelcref{LitimflowLPA} compared to the solutions to the Schr\"odinger equation~\labelcref{eq:RadialSchroedinger}.
Note that the finite-energy eigenvalue for $\theta=0$ includes the zero-point energy~\labelcref{eq:zeropoint energy}.
We see that the flow equation captures well the $\theta$-dependence of the energy eigenvalues.

\begin{figure}
    \centering
    \includegraphics[width=\linewidth]{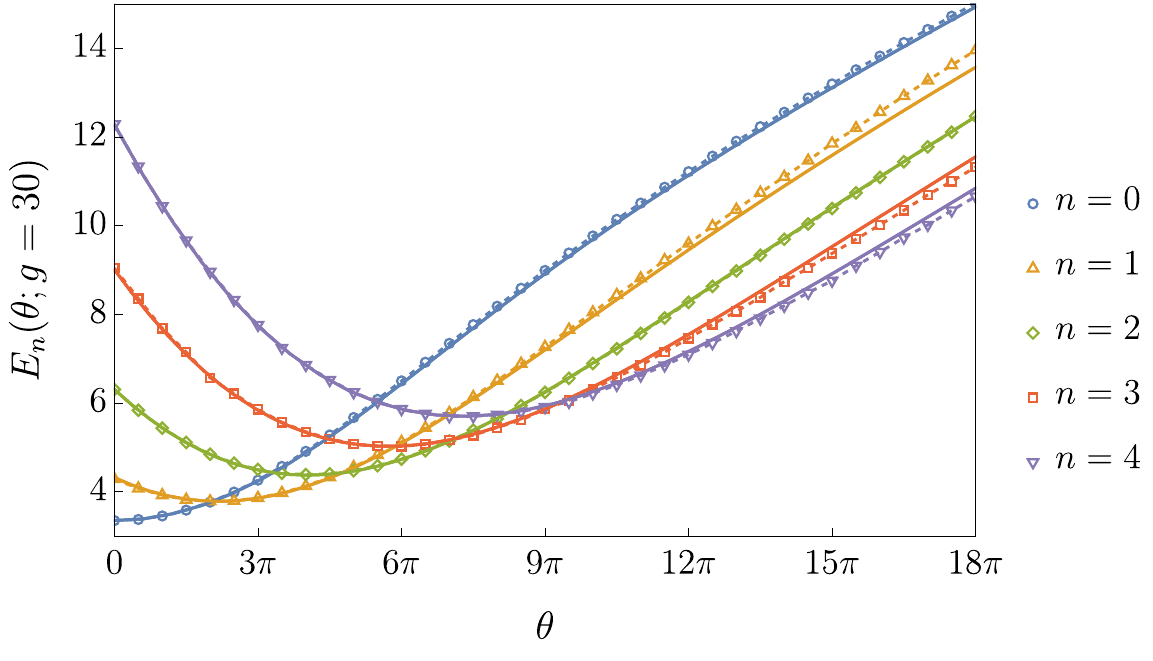}
    \caption{Energy levels~\labelcref{eq:Energy-fRG} for different $n$'s and $g=30$  in terms of the $\theta$-parameter from the flow equations with the Litim-type regulator. 
    The solid lines show the energy levels obtained from the Schr\"odinger equation~\labelcref{eq:RadialSchroedinger}.}
    \label{fig:FlowEnergy for diff n}
\end{figure}

For a fixed value $g=30$, each energy level ($n=0,\,1,\,2,\,3,\,4$) is displayed as a function of $\theta$ in \Cref{fig:FlowEnergy for diff n}.
One finds that the $\theta$-dependence is well captured for each quantum number $n$, or equivalently the winding number $\nu$.
However, in this approach, the level crossing in the ground-state energy is not observed because of the fixed topology (quantum number).

We close the discussion of the results for fixed topology with an (incomplete) assessment of the systematic error. For this purpose, we have also computed the energy levels with the CS-regulator. From the perspective of functional optimisation \cite{Litim:2000ci, Pawlowski:2005xe, Pawlowski:2015mlf}, this regulator is close to the 'worst', see \cite{Litim:2001hk}. The respective results are shown in \Cref{fig:n0energylevelCS} in \Cref{app:ResultsCS}. We see that the flow equation with the CS-regulator also captures the energy levels well, even though the Litim regulator leads to more accurate results than the CS-regulator as expected. This suggests a rather small systematic error of the present LPA results, but we hasten to add that the present considerations are far from a comprehensive systematic error analysis. 

%

%%%%%%%%%%%%%%%%%%%%%%%%%%%%%%%%%%%%
\subsection{Evolution of the effective potential with cusps}
\label{sec:EffPotCusps}

We proceed with the computation of the full effective potential $V(r^2,\theta)$ without any topological restrictions. This is done by starting the flow from the initial effective action in \labelcref{eq:ZFull}. For the sake of completeness we remark that this can be done directly for all $r$ and $\theta$ if we would monitor the position of the cusps during the flow: the flow in between the cusps is that of $V_n$ for a given $n$. In \Cref{app:CuspPostion} we show how this is done practically; see in particular \labelcref{eq:FlowCusp2,eq:cuspFlowFiniteVolume}. 
\begin{figure}
    \centering
    \includegraphics[width=0.8\linewidth]{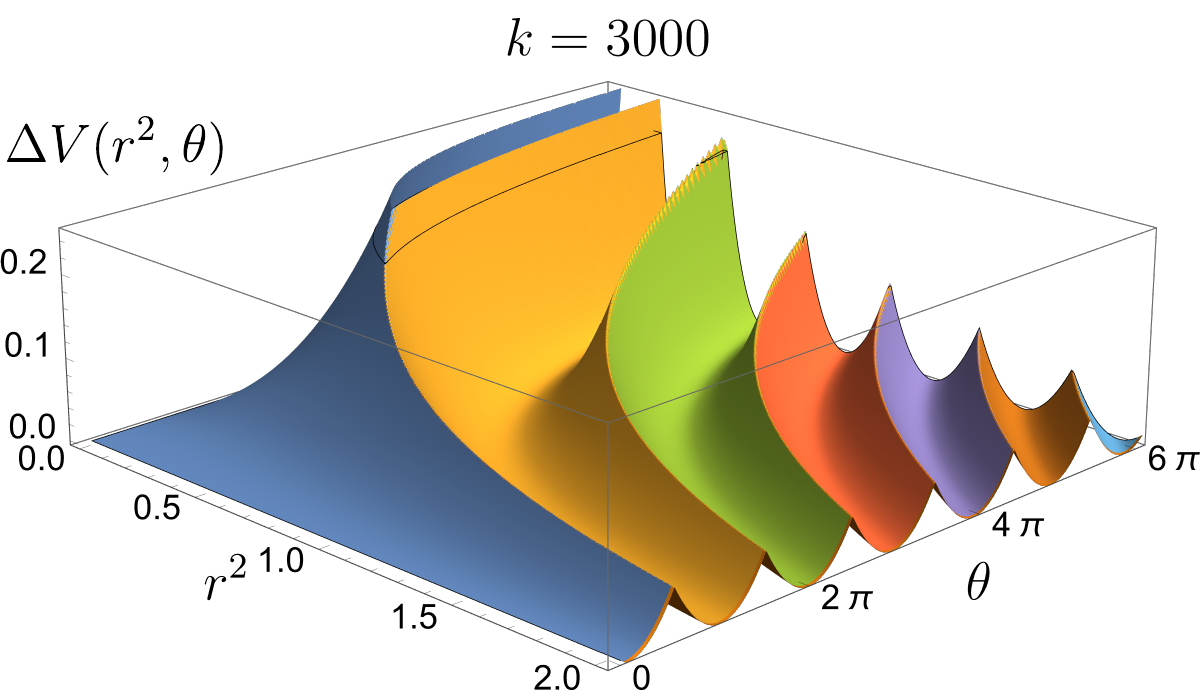}
    \includegraphics[width=0.8\linewidth]{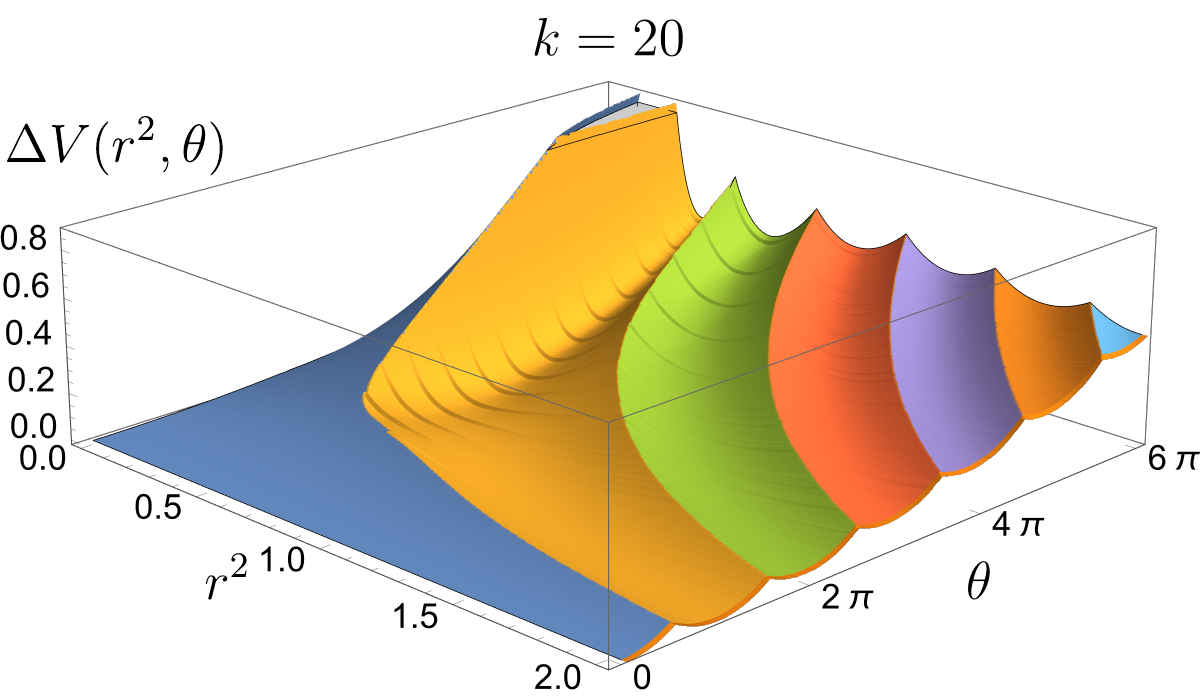}
    \includegraphics[width=0.8\linewidth]{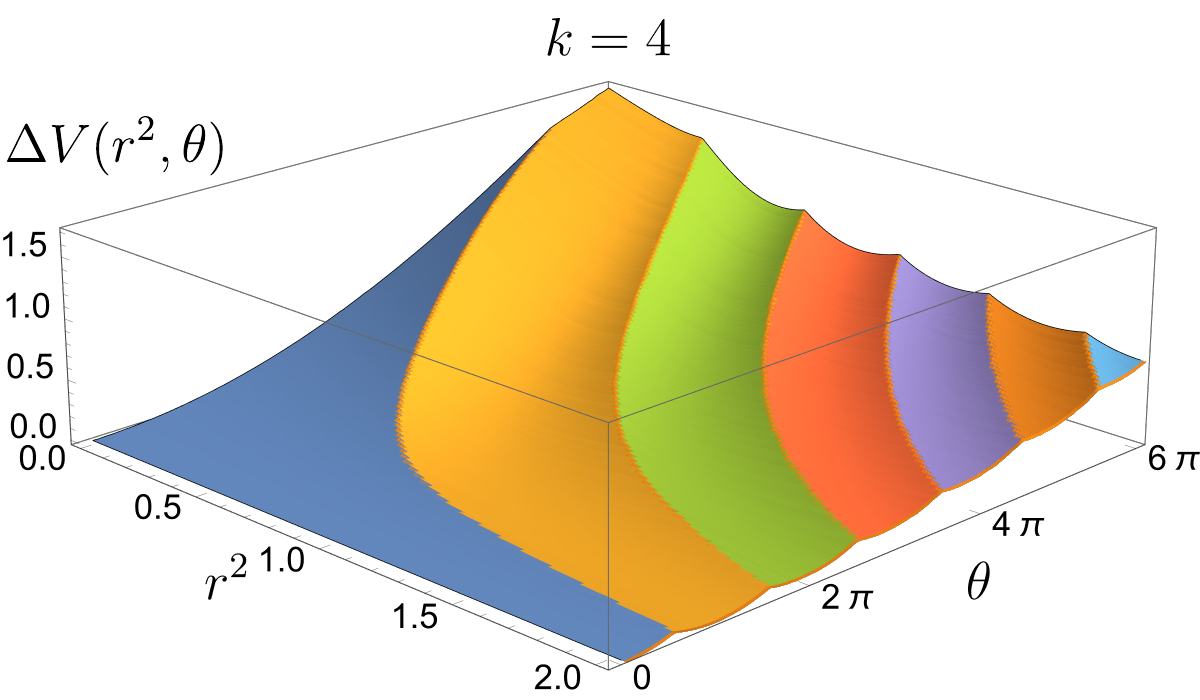}
    \includegraphics[width=0.8\linewidth]{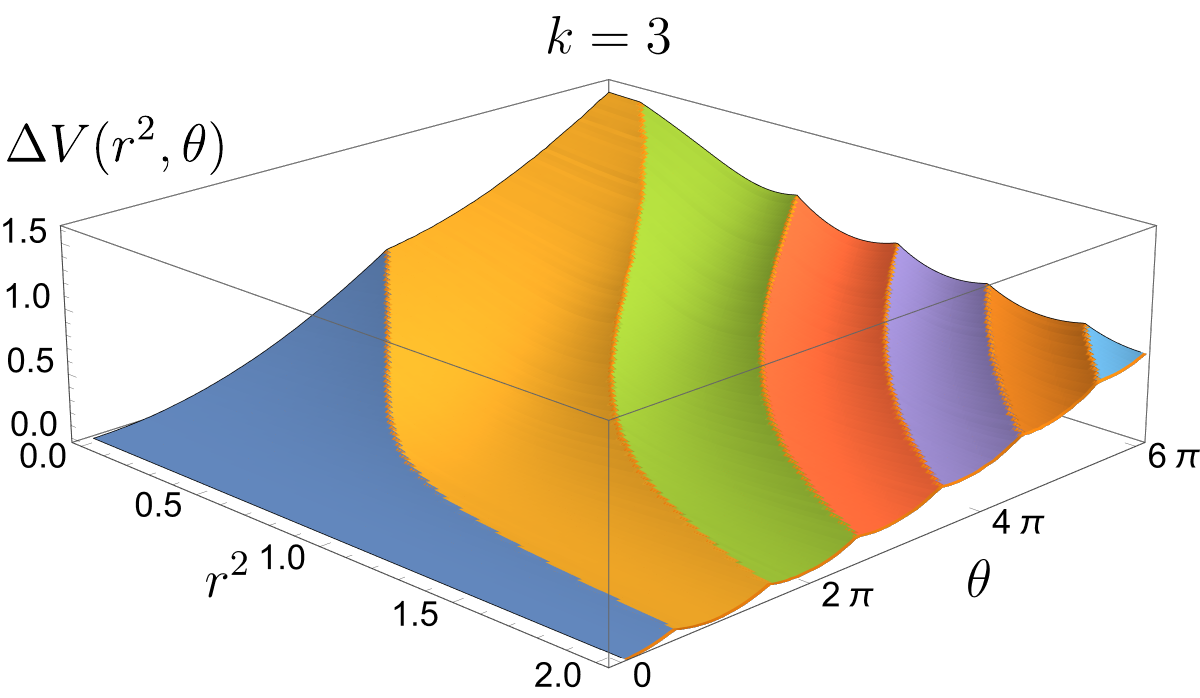}
    \includegraphics[width=0.8\linewidth]{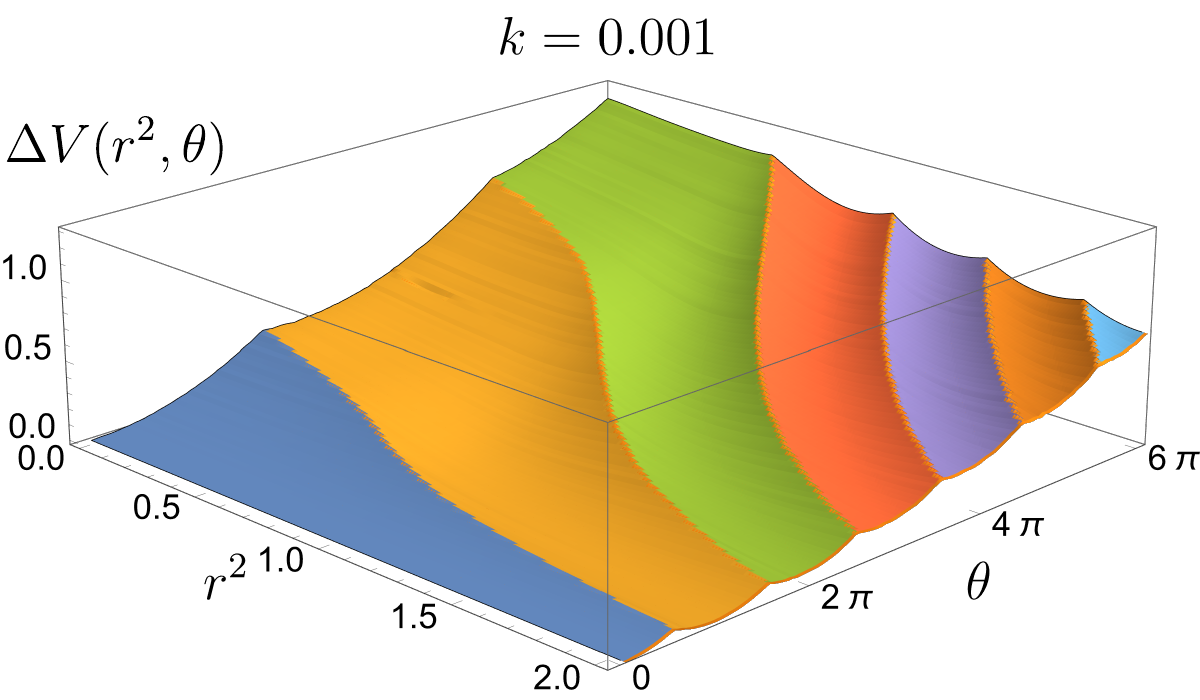}
    \caption{RG evolution of the effective potential \labelcref{eq:DeltaV} obtained from the flow equation with Litim-type regulator \labelcref{LitimflowLPA}.
    The different branches are characterised by different colors individually, showing the quantum number of the energy level. The bottom plot is the same as \Cref{fig:EffectivePotential}. Contour plot of the effective potential at each scale is depicted in \Cref{fig:jumpingFlowWithk}.}
    \label{fig:NonAnaPotential}
\end{figure}
However, while possible, the extraction of $V_k(r^2,\theta)$, does not necessitate new numerical computations, but simply the use of the effective potentials $V_n$ obtained in the different topological sectors: as mentioned above, within the cusps the flow of $V_k$ is simply that of $V_{n,k}$ for a given $n$. The flowing cusps are simply the intersection points (or level crossings) of the potentials $V_n$. 

\begin{figure}
    \centering
    \includegraphics[width=0.98\linewidth]{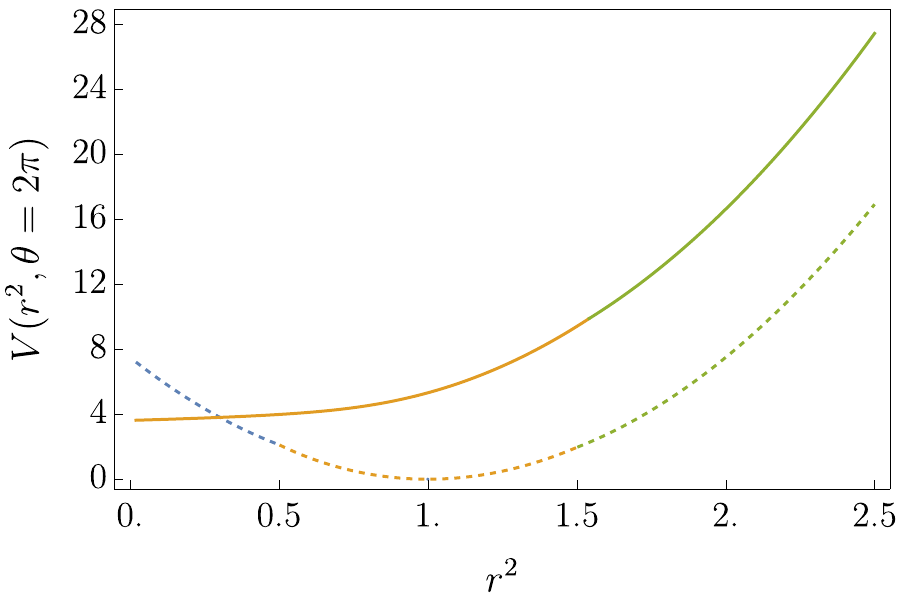}
    \caption{
    Effective potential $V(r^2,\theta=2\pi)$ evaluated at $k=3000$ (dashed) and $k=0.001$ (solid), from the flow equation with Litim-type regulator \labelcref{LitimflowLPA}. Different colors of the potential characterise the different branches of the quantum number $n$ defined in \labelcref{eq:initialConditionWithFixedn}.} \label{fig:PotentialFixedThetaNotSubtractedCusps}
\end{figure}

In the present work we are only interested in the potential at $k=0$. There we exploit \Cref{eq:Energy-fRG} which already includes the relative normalisation between the sectors. Then, the complete potential $V(r^2,\theta)$ is obtained by the following definition 
\begin{align}\nonumber 
V(r^2,\theta) = &\,\min_{n}\Bigl[ V_{n,k=0}(r^2,\theta)-V_{n,k=0}(\langle r\rangle^2,\theta) \\[1ex] 
&\hspace{1cm}+E_n^{\textrm{fRG}}(\theta;g)\Bigr]\,.
    \label{eq:Vfinal}
\end{align}
The respective full effective potential for $n\leq 7$ is shown in \Cref{fig:NonAnaPotential}, and carries the non-analytical crossing lines. For the sake of clarity we have subtracted the effective potential at $\theta = 0$ case, i.e., the plot shows 
\begin{align}
   \Delta  V(r^2, \theta) = V(r^2, \theta) - V(r^2, \theta = 0)\,.
    \label{eq:DeltaV}
\end{align}
We close the discussion of the full potential with an illustration of the non-triviality of the flow. For this purpose we plot the full effective potential with respect to the field variable $r^2$ for a fixed $\theta = 2\pi$, and different cutoff scales. We show $k$-snapshots at the initial scale $k = 3000$ and a deep infrared scale $k = 0.001$, where the flow has settled. The respective results are shown in  \Cref{fig:PotentialFixedThetaNotSubtractedCusps}, further results are presented in \Cref{app:RGevolutionofeffectivepotential}.

\begin{figure}
    \centering
\includegraphics[width=0.98\linewidth]{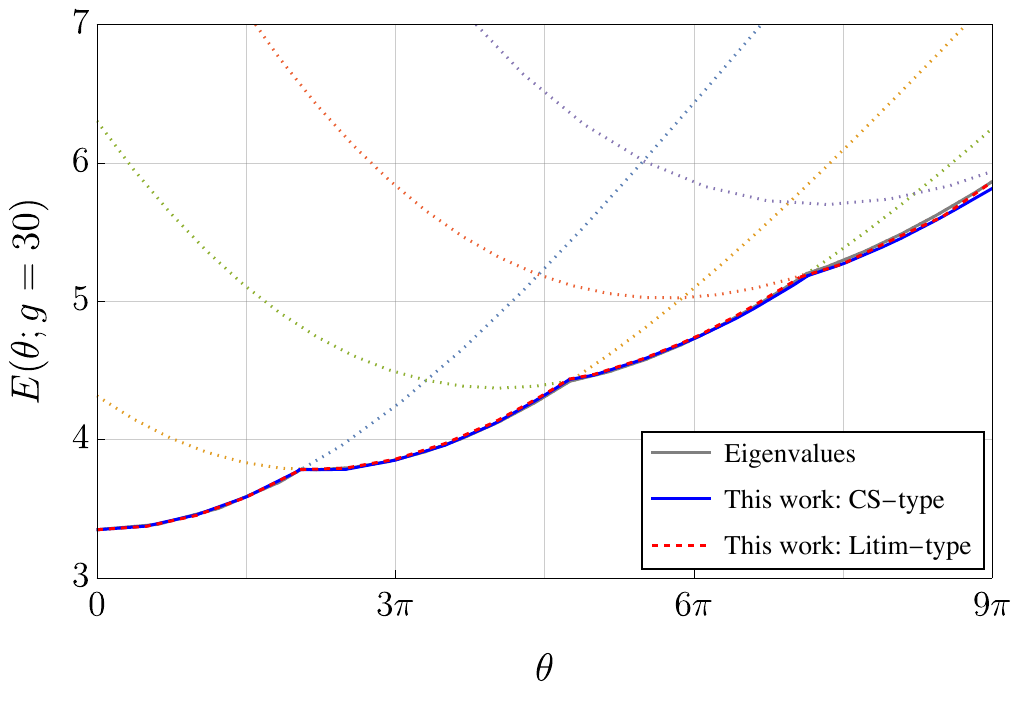}
\caption{Ground-state energy \labelcref{eq:Energy-fRGFinal}. We display results obtained with the flow equation with the CS-type regulator (dashed blue) and with the Litim-type regulator (dashed red). The solid gray line denotes the benchmark result obtained by solving the eigenvalue problem of the Schr\"odinger equation \labelcref{eq:RadialSchroedinger}. The colored dotted lines indicate the energies of the excited states, as shown in the right panel of \Cref{fig:thetaVacuumExact}.
    }  
   \label{fig:E-QM}
\end{figure}

\Cref{eq:Vfinal} leads us to the final expression for the energy of the system, 
\begin{align}
E(\theta;g) = V(r^2=\langle r\rangle^2)=
\min_{n} E_n^{\textrm{fRG}}(\theta;g)\,.
\label{eq:Energy-fRGFinal}
\end{align}
Within the current approximation, this result is displayed in \Cref{fig:E-QM} together with the exact numerical result. 

%

%%%%%%%%%%%%%%%%%%%%%%%%%%%%%%%%%%%%%%%%
\subsection{Topological susceptibility}
\label{sec:TopSuceptWith}

\begin{figure}[t!]
    \centering
 \includegraphics[width=0.98\linewidth]{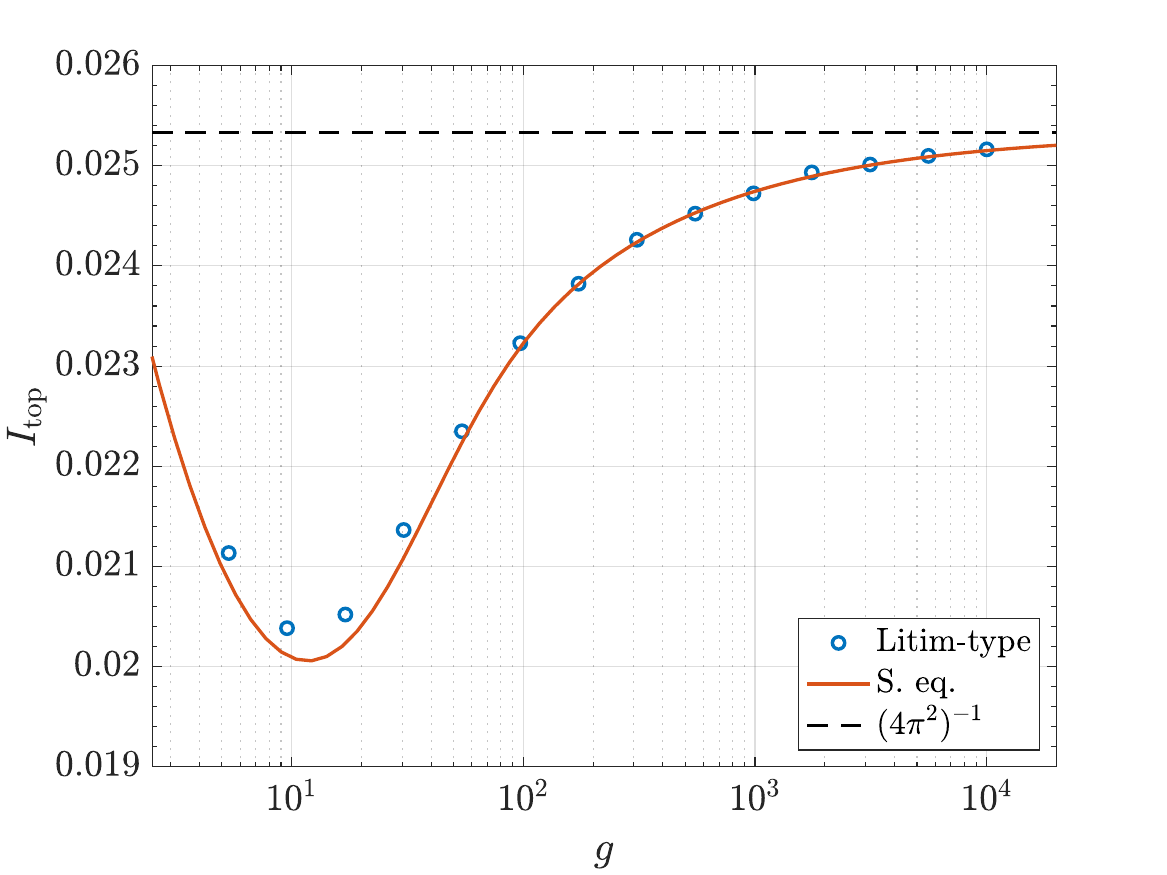}
    \caption{Topological susceptibility \labelcref{eq:TopI-g} with respect to the constraint coupling $g$. The LPA-fRG result (circles) is obtained by the Litim-type flow equation~\labelcref{LitimflowLPA}. The red line is the result from the Schr\"odinger equation~\labelcref{eq:RadialSchroedinger}. The dashed line exhibits the value~\labelcref{eq:TopI} in topological quantum mechanics.}
    \label{fig:Top_sus_alt}
\end{figure}

Finally, we extract the equal-time two-point correlation of the topological charge. We start this analysis with a brief discussion of the results obtained from the Schr\"odinger equation~\labelcref{eq:RadialSchroedinger}. The analytical result for $n=0$ in the  topological limit ($g\to \infty$) is given by
\begin{align}
    I_{\rm top} \equiv \braket{\dot{\varphi}(0)\dot{\varphi}(0)} = \frac{\partial^2}{\partial \theta^2} \ln Z(\theta) = \frac{1}{4 \pi^2}\,,
    \label{eq:TopI}
\end{align}
where we have used \labelcref{eq:EgInfty} with $m=1$. For a finite $g$, we evaluate
\begin{align}
I_{\rm top}(g) = \frac{\partial^2}{\partial \theta^2} E_0(\theta;g)\,.
\label{eq:TopI-g}
\end{align}
The results obtained from the Schr\"odinger equation are shown in \Cref{fig:Top_sus_alt}. For large $g\geq 1$, $I_{\rm top}(g)$ approaches \labelcref{eq:TopI}. Moreover, it features a minimum at $g = 11.75$. 
In \Cref{fig:Top_sus_alt} we compare the exact results with those obtained from the flow equation with the Litim regulator~\labelcref{LitimflowLPA}. 

For the extraction of the second derivatives of the ground state energy $E_0$ from the effective potential, we use the fact that in the zero-temperature limit, $\beta \rightarrow \infty$, the free energy is identical to the ground state energy itself. As for the effective potential itself, $I_{\rm top}(g)$ agrees well with the exact results. 

We close this Section with the remark that an obvious and interesting extension of the present analysis is the computation $\langle \dot\varphi(t)\dot\varphi(0) \rangle$. For this computation, one has to go beyond the LPA-approximation used in the present work, and the respective results will be presented elsewhere.

%

%%%%%%%%%%%%%%%%%%%%%%%%%%%%%%%%%%%%%%%%%
\section{Summary and Conclusion}
\label{sec:summary}

In the current work, we have studied $U(1)$-symmetric quantum mechanics with the functional renormalisation group. 
This analysis served as a benchmark test of the capability of the functional renormalisation group to incorporate topological properties of quantum systems. 

Specifically, we have computed the vacuum energy structure within the local potential approximation of the Wetterich flow. 
We have compared results obtained with different regulators, the Litim and Callan-Symanzik regulators, with the benchmarks obtained by directly solving the Schr\"odinger equation. 
The fRG results agree very well with the exact ones, showing an increasing, albeit small, error for larger $\theta$-parameters, related to large topological numbers $n$'s. 
In our opinion, this is a rather impressive agreement, given the qualitative nature of the LPA-approximation. 

We have also computed the topological susceptibility coefficient, which is also agreeing well with the exact results. 
Moreover, our relaxed system with $g<\infty$ reduces to the quantum rotor in the large coupling limit $g\rightarrow\infty$: 
our results show the correct asymptotic behaviour $I_{\rm top}\rightarrow 1/(2\pi)^{2}$ as $g\rightarrow\infty$, confirming the consistency of our framework with the expected topological properties of the model. 

The asymptotic matching underscores the robustness of our regularisation scheme in preserving topological information under renormalisation group flow, even at strong coupling. This is a critical prerequisite for studying systems like gauge theories where topology is inherently tied to non-perturbative dynamics. 

In summary, our work demonstrates the potential of fRG to address topological aspects of quantum systems in a well-controlled but non-trivial benchmark theory. Our results complement those in the anharmonic oscillator in \cite{Bonanno:2025mon}. The analysis in the present work also provides a road map for the application of our findings and techniques to the $\theta$-vacuum structure in gauge theories. While we specifically aim at QCD and respective computations are under way, a first step in this programme is provided by the study of the 2D Schwinger model. We hope to report on results for both cases soon.

%%%%%%%%%%%%%%%%%%%%%%%%%
\begin{acknowledgments}

We thank Akio Tomiya for discussions.
%%%%%%
YG and MY acknowledge the Institute for Theoretical Physics, Heidelberg University, and Tokyo Women's Christian University for the very kind hospitality during their stay.
%%%%%%
This work was supported by the National Science Foundation of China (NSFC) under Grant No.~12205116 and the Seeds Funding of Jilin University. 
%%%%%%%%%%
JMP is funded by the Deutsche Forschungsgemeinschaft (DFG, German Research Foundation) under Germany’s Excellence Strategy EXC 2181/1 - 390900948 (the Heidelberg STRUCTURES Excellence Cluster) and the Collaborative Research Centre SFB 1225 - 273811115 (ISOQUANT).
\end{acknowledgments}

\onecolumngrid
\appendix
%%%%%%%%%%%%%%%%%%%%%%%%%%%%%%%%%%%%%%%%%
\section{Quantum mechanics with topological term}
\label{app:QMwithTopTerm}

In this Appendix, we summarise the basic properties of quantum mechanics on $S^1$, the quantum rotor and its relaxed version. In \Cref{app:QuantumRotor} we briefly discuss the quantum rotor with field variables $z\in U(1)$. In 
\Cref{app:RelaxedQuantumRotor} we relax the $U(1)$-constraint and consider $z\to \varphi\in \mathbbm{C}$.

%%%%%%%%%%%%%%%%%%%%%%%%%%%%%%%%%%%%%%%%%%
\subsection{Quantum rotor and its topological vacuum structure}
\label{app:QuantumRotor}

The quantum rotor is defined with the classical action 
\begin{align}
    S[z] =\int_\tau \left[ \frac{m}{2} \dot{z}^* \dot{z} - \frac{\theta}{4\pi} (z^* \dot{z} - \dot{z}^* z) \right]\,,
    \label{Appeq:U1_QM_action}
\end{align}
where the dot represents the derivative with respect to $\tau$. The field variable $z = e^{i \vartheta}$ is an element of $U(1)$. 
As shown in \labelcref{eq:winding_number}, the second term in \labelcref{Appeq:U1_QM_action} corresponds to a topological term. The path integral for this system is given by
\begin{align}
   Z_z[J_z] = \int \mathcal{D} \hat z  \, e^{-S_\theta[z] + \int_\tau \, J \hat z}\,,
    \label{eq:ZQuantumRotor}
\end{align}
where $\mathcal{D} z$ is the Haar measure of $U(1)$. The topological term does not affect the classical equations of motion for $z$; however it contributes to the path integral through the summation over field configurations with different boundary conditions~\labelcref{eq:phi_boundary_condition} which corresponds to $z_\nu(\beta) = z_\nu(0) \cdot e^{2 \pi i \nu}$. 
To elucidate this, we define the field configurations $\{ z_\nu(\tau) \}$ as
\begin{align}
    z_\nu(\tau) = e^{i \frac{2 \pi \nu}{\beta} \tau} z_0(\tau)\,, \quad \text{with} \quad z_0(\tau + \beta) = z_0(\tau)\,.
\end{align}
This corresponds to a local gauge transformation where the phase transforms as $\vartheta \to \vartheta + \delta \vartheta_\nu$ with $\delta \vartheta_\nu = \frac{2 \pi \nu}{\beta} \tau$.

For vanishing sources, the path integral reduces to 
\begin{align}
    \mathcal{Z} = \sum_{\nu \in \mathbb{Z}} \int [\mathcal{D} z_\nu] \, e^{-S[z_\nu(\tau), z^*_\nu(\tau)]}
    = \mathcal{N}(\beta) \sum_{\nu \in \mathbb{Z}} e^{-\frac{2\pi^2 m}{\beta} \nu^2 + i \theta \nu}\,,
    \label{eq:winding_sum}
\end{align}
where $\mathcal{N}(\beta)$ is a factor independent of $\theta$ and $w$, defined as
\begin{align}
    \mathcal{N}(\beta) \equiv \int [\mathcal{D} z_0] \, e^{-\int_0^\beta \mathrm{d}\tau \, L}
    =\sqrt{\frac{\beta}{2\pi m}}\,,
\end{align}
carrying the information of kinetic fluctuations.
Applying the Poisson resummation formula,
\begin{align}
    \sum_{\nu \in \mathbb{Z}} s(\nu) = \sum_{n \in \mathbb{Z}} \int_{-\infty}^{+\infty} \mathrm{d}x \, s(x) e^{i 2 \pi n x} \equiv \sum_{n \in \mathbb{Z}} S(n)\,,
\end{align}
to \labelcref{eq:winding_sum}, we obtain
\begin{align}
    \mathcal{Z} = \mathcal{N}'(\beta) \sum_{n \in \mathbb{Z}} e^{-\frac{\beta}{2m} \left(n - \frac{\theta}{2\pi}\right)^2}\,.
\end{align}
This result takes the form of the partition function for a canonical ensemble, and thus the energy eigenvalues are found to be
\begin{align}
    E_n = \frac{1}{2m} \left(n - \frac{\theta}{2\pi}\right)^2\,.
\end{align}
Here, $n \in \mathbb{Z}$ is the quantum number labeling the energy spectrum, arising from the quantisation of angular fluctuations.

%%%%%%%%%%%%%%%%%%%%%%%%%%%%%%%%%%%%%%
\subsection{Relaxation to a \texorpdfstring{$U(1)$}{}-symmetric theory}
\label{app:RelaxedQuantumRotor}

The topological nature \labelcref{eq:winding_number} arises from the field constraint $z^*z=1$.
Now, we relax this condition, 
\begin{align}
    z = e^{i \vartheta} \longrightarrow \varphi = r \,e^{i\vartheta}\,,
\end{align}
and deal with $\varphi$ and $\varphi^*$ as quantum mechanical variables.
We have $\varphi^* \varphi =r^2\in \mathbbm{R}^+$ in general, and hence $\varphi \in \mathbbm{C}$. 
The respective classical action reads
\begin{align} 
    &S[\varphi,\varphi^*] =  
    \int_\tau \, \left[ \frac{m}{2} \dot{\varphi}^* \dot{\varphi} - \frac{\theta}{4\pi} (\varphi^* \dot{\varphi} - \dot{\varphi}^* \varphi) + V(\varphi^*\varphi) \right]\,,
\label{eq:SvarphiApp}
\end{align}
with the potential 
\begin{align}
    V(\varphi^*\varphi) = V(r^2)= \frac{g}{4} (\varphi^*\varphi - 1)^2 
     = \frac{g}{4} (r^2-1)^2\,,
\label{eq:Vvarphir} 
\end{align}
with the global minimum at $\varphi^*\varphi =r^2 = 1$. The model reduces to the quantum rotor for $g\to \infty$. Then, the potential simply introduces the constraint $\varphi\varphi^*=1$. The path integral for this system is given by 
\begin{align}
    \mathcal{Z}[J_\varphi] = \int \mathcal{D}\hat\varphi \, e^{-S_\theta[\varphi,\varphi^*] + \int_\tau \, J_\varphi \hat\varphi}\,,
    \label{eq:ZRelaxedQuantumRotor}
\end{align}
see also \labelcref{eq:ZFull} with $\hat\phi=\hat\varphi$. 
For finite coupling $g$, the quantum rotor with the generating functional $Z_z$ in \labelcref{eq:ZQuantumRotor} is  deformed into a $U(1)$-symmetric model with the complex field $\varphi$. In terms of the fields $r$ and $\vartheta$, the classical action \labelcref{eq:SvarphiApp} is given by
\begin{align}
    S[r,\vartheta]= \int_\tau\left[\frac{m}{2} ( \dot{r}^2 + r^2 \dot{\vartheta}^2 ) + \frac{\theta}{2\pi} r^2 \dot{\vartheta} - \frac{g}{4} (r^2 - 1)^2\right]\,.
    \label{eq:SvarphirApp}
\end{align}
The respective Hamiltonian follows with a Legendre transformation from the Lagrangian in \labelcref{eq:SvarphirApp} as 
\begin{align}
&H
=\frac{1}{2 m}\Pi_r^2
+ \frac{1}{2 m r^2} \left( \Pi_\vartheta - \frac{\theta}{2 \pi} r^2 \right)^2 + \frac{g}{4} (r^2 - 1)^2\,,
\end{align}
where we have defined the canonical momenta
\begin{align}
    \Pi_r = m \dot{r}, \qquad 
    \Pi_\vartheta = m r^2 \dot{\vartheta} + \frac{\theta}{2\pi}r^2\,.
\end{align}
With canonical quantisation, the  canonical momentum operators in polar coordinates are given by
\begin{align}
   \hat \Pi_r^2 = -\frac{1}{r}\frac{\partial}{\partial r}\left(r \frac{\partial}{\partial r} \right)\,, \qquad \qquad 
   \hat \Pi_\vartheta = -i \frac{\partial}{\partial \vartheta}\,, 
\end{align}
and the time-independent Schr\"odinger equation is given by \labelcref{eq:RadialSchroedinger}.

%%%%%%%%%%%%%%%%%%%%%%%%%%%%%%%%%%%%%%%%%%%%%%%%%%%%%%%%%%%%%%%%%%%%%
\section{Derivation of the flow equation in the local potential approximation}
\label{sec:DerivationFlowApp}

The effective action in the local potential approximation (LPA) is given by
\begin{align}
    \Gamma_k[\varphi] 
    = \int \frac{\df p}{2\pi} \left[ \frac{m}{2} \varphi^*(-p) p^2 \varphi(p)\right] + \int_\tau V_k(\varphi^*\varphi) \,,
    \label{appeq:GammakWithTopResum}
\end{align}
where the potential $V_k$ is the only cutoff-dependent part in the effective action, and $\partial_t m=0$. Within a slight abuse of notation we use $\varphi$ for both $\varphi(\tau)$ and its Fourier transformation $\varphi(p)$ with 
\begin{align}
     \varphi(\tau) = \int \frac{\df p}{2\pi} \varphi(p) e^{-i p \tau}\,.
\end{align}
We proceed with the derivation of the flow equation for the effective action \labelcref{eq:GammakWithTopResum} and a generic regulator term for the complex field $\varphi$, 
\begin{align} 
\Delta S_k[\varphi] = \frac{m}{2} \int_p \varphi^*(-p)\,R_k(p) \,\varphi(p)\,.
\label{eq:DeltaSku}
\end{align}
In the $(\varphi,\varphi^*)$-basis, the two-point function reads 
\begin{align}
     &\Gamma_{k}^{(2)}(p,q)= 2\pi\, \delta(p+q)
\left(
    \begin{array}{cc}
        V_{\varphi\varphi} & \frac{m}{2} p^2  + V_{\varphi\varphi^*} \\[1ex]
        \frac{m}{2} p^2 + V_{\varphi^*\varphi} & V_{\varphi^*\varphi^*}
    \end{array}
    \right)\,,
    \label{eq:Gammudaggeru}
\end{align}
with 
\begin{align}
&V_{\varphi} \equiv \partial^2_{\varphi} V_k =\varphi^{*2} V_k^{\prime\prime}\,, \qquad V_{\varphi^*\varphi^*} \equiv \p_{\varphi^*}^2V_k = \varphi^{2} V_k^{\prime\prime}\,,\qquad V_{\varphi^* \varphi} = V_{\varphi\varphi^*}\equiv \partial_{\varphi}\p_{\varphi^*} V_k =V_k^{\prime} +r^{2} V_k^{\prime\prime}\,, 
\end{align}
with $V_k^\prime(x) = \partial_x V_x(x)$ and $V_k^{\prime\prime}(x) = \partial^2_x V_x(x)$. The regulator matrix $R_k^{ab}(p,q)$ is given by 
\begin{align}
  (R^{ab}_k)=2\pi\, \delta(p+q)\pmat{
      0   & \frac{m}{2}{R}_k(p) \\[1ex]
        \frac{m}{2}{R}_k(p) & 0
}\,, 
\end{align}
where the superscript $^{ab}$ now indicates $\varphi$ and $\varphi^*$. 
With these preparations, the right-hand side of the flow equation~\labelcref{eq:wettericheq} follows as 
\begin{align}\nonumber 
\frac{1}{2}\Tr\left[\frac{1}{\Gamma_k^{(2)} +R_k} \p_t R_k \right]& \\[2ex]\nonumber 
&\hspace{-2cm} =\frac{1}{2}\Tr\left[\left(
    \begin{array}{cc}
       \varphi^{*2} V_k^{\prime\prime} & \frac{m}{2} P_k(p)  + V_k^{\prime} +r^{2} V_k^{\prime\prime} \\[2ex]
        \frac{m}{2} P_k(p) + V_k^{\prime} +r^{2} V_k^{\prime\prime} & \varphi^{2} V_k^{\prime\prime}
    \end{array}
    \right)^{-1}\pmat{
      0   & \frac{m}{2}\p_t {R}_k(p) \\[1ex]
        \frac{m}{2}\p_t{R}_k(p) & 0
} \,2\pi\,\delta(p+q) \right]\\[2ex]
&\hspace{-2cm} = {\cal V}_\tau \int \frac{\df p}{2\pi}\,\frac{  \frac{m}{2}P_k(p) + V^{\prime} + r^2 V^{\prime\prime} }{\left[ \frac{m}{2} P_k(p) + V^{\prime} + r^2 V^{\prime\prime} \right]^2 - \left[ r^2 V^{\prime\prime} \right]^2}\,\frac{m}{2}\partial_t {R}_k(p)\,,
\end{align}
where 
\begin{align} 
P_k(p)=p^2 + R_k(p)\,, \qquad r^2=\varphi^*\varphi\,,\qquad  {\cal V}_\tau= \int \textrm{d}\tau = 2\pi\,\delta(0)\,, 
\end{align} 
Then, the flow equation for the effective potential is given by
\begin{align}
\partial_t V_k
    = \frac{1}{{\cal V}_\tau}\p_t\Gamma_k
    &= \int \frac{\df p}{2\pi} \frac{ \frac{m}{2}P_k(p) + V^{\prime} + r^2 V^{\prime\prime} }{\left[ \frac{m}{2} P_k(p) + V^{\prime} + r^2 V^{\prime\prime} \right]^2 - \left[ r^2 V^{\prime\prime} \right]^2}\,\frac{m}{2}\partial_t {R}_k(p)\,.
\label{eq:dtVApp}
\end{align}
For the Litim-type regulator \labelcref{eq:LitimReg} the derivative of the regulator with respect to the dimensionless renormalisation scale is obtained as
\begin{align}
    \partial_t R_k = 2 k^2 \Theta\left(k^2 - p^2\right) + 2 k^2 \left(k^2 - p^2\right) \delta\left(k^2 - p^2\right)= 2 k^2 \Theta\left(k^2 - p^2\right)\,.
\label{eq:LitimRegDer}
\end{align}
The flow equation of the effective potential with the Litim-type regulator follows from \labelcref{eq:dtVApp,eq:LitimRegDer} as 
\begin{align}\nonumber 
    \partial_t V_k(r^2) &= \int \frac{\df p}{2\pi}\left[ \frac{ \frac{m}{2} p^2 + \frac{m}{2} R_k + V^{\prime}(r^2) + r^2 V^{\prime\prime}(r^2) }{\left[ \frac{m}{2} p^2 + \frac{m}{2} R_k + V^{\prime}(r^2) + r^2 V^{\prime\prime}(r^2) \right]^2 - \left[ r^2 V^{\prime\prime}(r^2) \right]^2}\,m\, k^2\,\right] \Theta\left( k^2 - p^2\right)\\[2ex]
    &= \frac{2mk^3}{\pi} \frac{m k^2 + 2V^{\prime}(r^2) + 2r^2 V^{\prime\prime}(r^2)}{\left[ m k^2 + 2 V^{\prime}(r^2) + 2 r^2 V^{\prime\prime}(r^2) \right]^2 - \left[ 2 r^2 V^{\prime\prime}(r^2) \right]^2}\,.
\label{eq:FlowVApp}
\end{align}
\Cref{eq:FlowVApp}, see also \labelcref{LitimflowLPA} in \Cref{sec:fRG_setup}, is used for the explicit computations in \Cref{sec:NumericalResults}.

%%%%%%%%%%%%%%%%%%%%%%%%%%%%%%%%%%%%%%%%%%%%%%%%%%
\section{Reference computations with the CS-regulator}
\label{app:ResultsCS}

\begin{figure}
    \centering
    \includegraphics[width=0.43\linewidth]{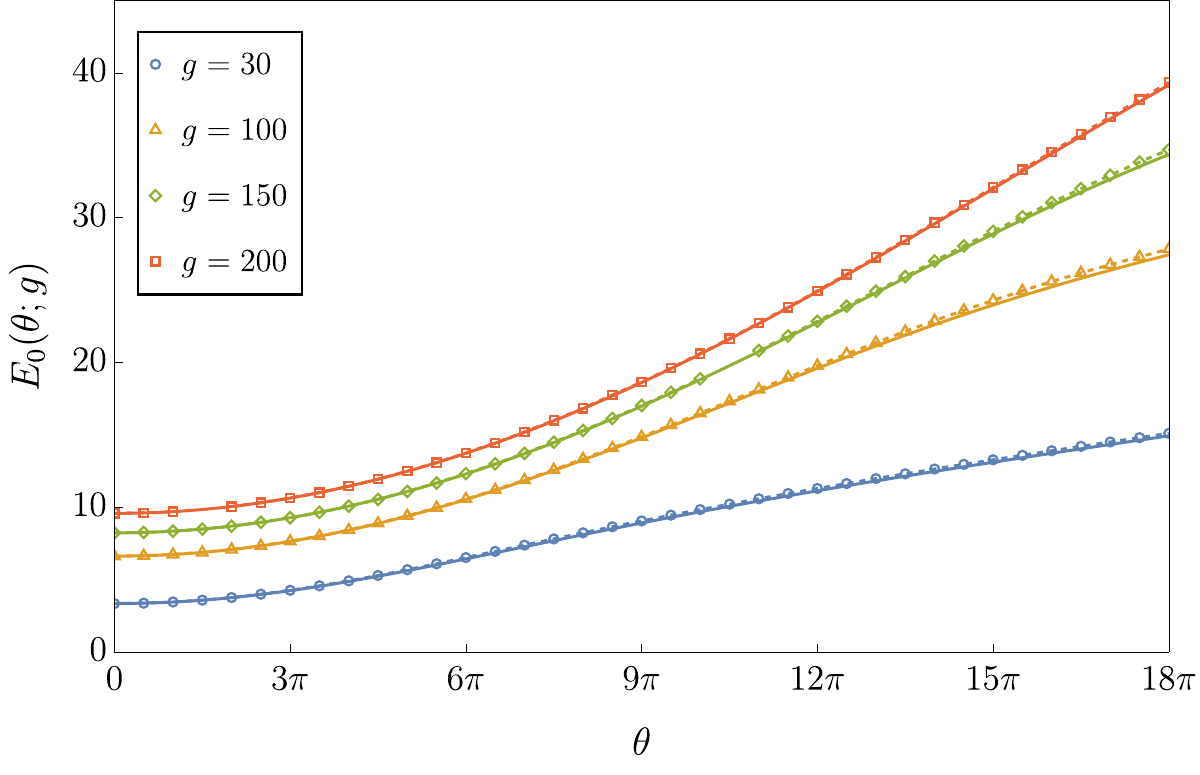}  
    \hspace{4ex}
    \includegraphics[width=0.5\linewidth]{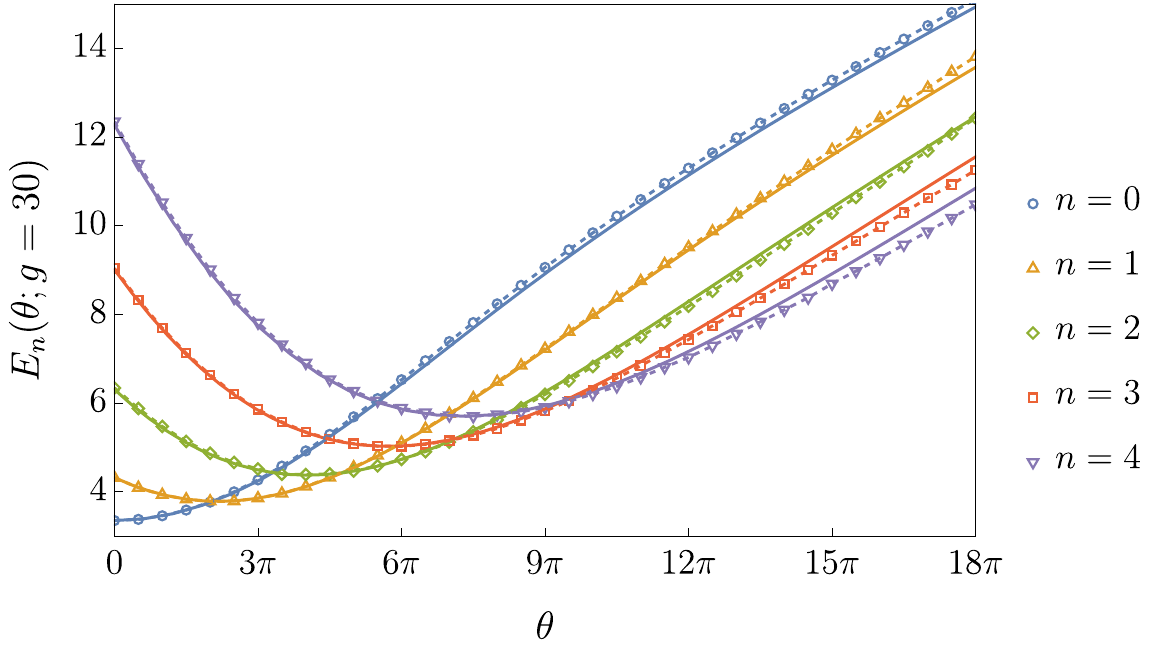}
    \caption{Energy eigenvalues $E_n$ (markers) as functions of the $\theta$-parameter obtained from the flow equations with the CS regulator~\labelcref{eq:floweqCS}. The left panel shows $E_0$ for various values of $g$, while the right panel displays $E_n$ for several values of $n$ at fixed $g=30$.
    The solid lines show the solution of the Schr\"odinger equation~\labelcref{eq:RadialSchroedinger}.}
    \label{fig:n0energylevelCS}
\end{figure}

We also provide reference computations for a CS-type regulator. We substitute the Litim or flat regulator function  \labelcref{eq:LitimReg} with 
\begin{align}
    R_k(p) = k^2\,.
    \label{eq:CSReg}
\end{align}
In view of optimisation this regulator is far away from the optimal one, see e.g.~\cite{Litim:2001hk}. However, in terms of analyticity in frequency space, related to causality, it is a very attractive choice, see \cite{Braun:2022mgx}. For the relaxed $U(1)$ model, the flow equation of the effective potential reads
\begin{align}\nonumber 
    \partial_t V_k =&\, \int_{-\infty}^{\infty} \frac{dp}{2\pi} \frac{mk^2\left[ \frac{m}{2} (p^2 + k^2) + V^\prime + r^2 V^{\prime\prime} \right]}{\left[ \frac{m}{2} (p^2 + k^2) + V^\prime + r^2 V^{\prime\prime} \right]^2 - \left[ r^2 V^{\prime\prime} \right]^2}\\[2ex]\nonumber 
    =& \,\int_{-\infty}^{\infty} \frac{dp}{2\pi} \frac{2 k^2\left[ (p^2 + k^2) + (2 V^\prime + 2 r^2 V^{\prime\prime})/m \right]}{\left[ p^2 + k^2 + 2V^\prime/m \right] \left[ p^2 + k^2 + (2V^\prime + 4 r^2 V^{\prime\prime})/m \right]} \\[2ex]
    =& \, \int_{-\infty}^{\infty} \frac{dp}{2\pi} \frac{k^2\left[ 2 p^2 + c_k^{(1)} + c_k^{(2)} \right]}{\left( p + i \sqrt{c_k^{(1)}} \right)\left( p - i \sqrt{c_k^{(1)}} \right)\left( p + i \sqrt{c_k^{(2)}} \right)\left( p - i \sqrt{c_k^{(2)}} \right)}\,, 
\label{eq:CSFlow}
\end{align} 
with 
\begin{align}
    &c^{(1)}_k = k^2 + 2 V^{\prime}(r^2)/m\,, &
    &c^{(2)}_k = k^2 + 2 V^{\prime}(r^2)/m + 4 r^2 V^{\prime\prime}(r)/m\,.
    \label{eq:C1andC2}
\end{align}
The integral can be performed analytically, which leads us to 
\begin{align}
    \partial_t V_k = \frac{k^2}{2}\left[ \frac{1}{\sqrt{c^{(1)}_k}} + \frac{1}{\sqrt{c^{(2)}_k}} \right]\,.
    \label{eq:floweqCS}
\end{align}
We have solved the flow equation with the CS-regulator \labelcref{eq:floweqCS} and computed the energy levels  \labelcref{eq:Energy-fRG}.
The left-hand side panel of \Cref{fig:n0energylevelCS} depicts the ground-state energy $E_0(\theta;g)$ with several values of $g$ as a function of $\theta$.
In the right-hand side panel of \Cref{fig:n0energylevelCS}, we also show the $\theta$-dependence of the energy levels ($n=0,\ 1,\ 2,\ 3,\ 4$) with $g=30$. 

The results with the CS-regulator in \Cref{fig:n0energylevelCS} have to be compared with that obtained with the Litim regulator in \Cref{fig:n0energylevel} and \Cref{fig:FlowEnergy for diff n}. In particular for the higher energy levels, \Cref{fig:FlowEnergy for diff n} and right-hand side panel of \Cref{fig:n0energylevelCS}, the increasing deviation from the exact results is more pronounced for the CS-regulator. 

%%%%%%%%%%%%%%%%%%%%%%%%%%%%%%%%%%
%%%%%%%%%%%%%%%%%%%%%%%%%%%%%%%%%%

%%%%%%%%%%%%%%%%%%%%%%%%%%%%%%%%%%%%%%%%%%%%%%%%%%
\section{Breakdown of level crossing capture}
\label{sec:BreakdownLC}

In this Appendix we derive the flow equation without topological resummation and demonstrate that such a setup fails to capture the level crossing of the ground-state energy. We resort to LPA, 
\begin{align}
\Gamma_k = \int_\tau \, \left[ \frac{m}{2} \dot{\varphi}^* \dot{\varphi} - \frac{\theta}{4\pi} (\varphi^* \dot{\varphi} - \dot{\varphi}^* \varphi) + V_k(\varphi^* \varphi) \right]\,,
\label{eq:EffectivAactionWithoutTop}
\end{align}
where the initial effective potential $V_\Lambda$ is given by \labelcref{eq:originalInitialPotential}. In contradistinction to the fRG approach put forward in the present work, the topological term is included in the effective action instead of being summed over. We expect that then 'topological freezing' will manifest itself in a regulator- and setup-dependent way. Below we discuss two different setups and assess their shortcomings: In \Cref{sec:FlowEquationInRealVariableBasis} we follow \cite{Fukushima:2022zor} and simply use a standard Cartesian regulator. In \Cref{subsec:BreakdownLevelCrossing} we make full use of the complex frequency setup and its similarity to the Silver Blaze properties in quantum field theories with a chemical potential.

%%%%%%%%%%%%%%%%%%%%%%%%%%%%%%%%%%%%%%
\subsection{Flow with topological freezing}
\label{sec:FlowEquationInRealVariableBasis}

In this Appendix we briefly review the fRG approach used in~\cite{Fukushima:2022zor}. There, the flow equation for the  effective action in \labelcref{eq:EffectivAactionWithoutTop} was derived analogously to \Cref{sec:DerivationFlowApp}. However, instead of the complex field basis a Cartesian basis was used. The field $\varphi$ was disentangled in its real and imaginary part, $\varphi = x + i y$.  
Using the Litim cutoff function \labelcref{eq:LitimReg}, the flow equation for the effective potential takes the form  
\begin{align}
\partial_t V_k = \frac{2mk}{|\theta|} \cdot \frac{mk^2 + 2V' + 2r^2 V''}{\sqrt{(mk^2 + 2V' + 2r^2 V'')^2 - (2r^2 V'')^2}} \arctan\left( \frac{k|\theta|}{\pi \sqrt{(mk^2 + 2V' + 2r^2 V'')^2 - (2r^2 V'')^2}} \right)\,.
\label{eq:FlowVk+theta-term}
\end{align}
A detailed derivation of this flow equation is provided in~\cite{Fukushima:2022zor}, and the only difference to \Cref{sec:DerivationFlowApp} is the topological term in the effective action. 

%Now we use the numerical solution of \labelcref{eq:FlowVk+theta-term} with the initial potential \labelcref{eq:originalInitialPotential} for studying the $\theta$-dependence of the ground-state energy. For a benchmark coupling value of $g = 30$, the flow equation yields results for the ground-state energy that closely match those obtained from the Schr\"odinger equation within the regime $\theta \lesssim 2\pi$, where $n = 0$ corresponds to the ground state.  
%However, as $\theta$ increases beyond $2\pi$, the solution to the flow equation breaks down due to the emergence of a pole in the flow, leading to a termination of the flow of the ground-state energy around $\theta \sim 2\pi$. This is also seeing in Figure~4 in~\cite{Fukushima:2022zor} or the left-hand side panel in \Cref{fig:LitimRealFlowEnergy}. 

For the numerical solution of \labelcref{eq:FlowVk+theta-term} it is more convenient to consider the  flow of the derivative of the potential $V_k^\prime(r^2) \equiv \partial_{r^2}V_k(r^2)$ rather than the potential $V_k(r^2)$ itself: The $r^2$-derivative leads to higher order poles in the flow equation for $\Gamma^{(2)}+R_k\to 0$. These poles are repulsive and encode the convexity-restoring property of the Wetterich equation, see \cite{Litim:2006nn}. While soft or even integrable singularities in the flow equation still accommodate convexity, the numerical access is far more challenging.   

Consequently we consider 
\begin{align}
    \partial_t V_k^\prime(r^2) = \frac{\partial}{\partial r^2} \Big[ \partial_t V_k(r^2) \Big]\,. 
\end{align}
This allows us to compute $V^\prime_{k=0}$ and the effective potential is obtained from an $r^2$-integration, 
\begin{align}
    V_{k}(r^2) = \int_{r_0^2}^{r^2} \df (r_1^2) \ V_k^\prime(r_1^2)+\textrm{const.}\,,
\end{align}
where the constant includes a potential shift such as used in \Cref{sec:EffPotCusps,sec:EnergyLevelsWithout}. 

In the left-hand side panel of \Cref{fig:LitimRealFlowEnergy}, we show the result following the above strategy in the red line with markers, with $r_0^2 = 1$ as the boundary field value. Evidently, the breakdown of the numerical solution of the flow at $\theta = 2\pi$, reported on in \cite{Fukushima:2022zor}, is absent. Indeed, this breakdown can be traced back to the soft singularity structure of the flow $\partial_t V$ which is resolved by considering $\partial_t V^\prime$. 

The result for $V(r)$ does not show the level crossing in line with the arguments in \Cref{sec:Initial+Top}: the flows do not accommodate topological configurations and hence the level crossing is absent. However the derivation still includes the topological term. It is suggestive that this term deforms the correct flow which leads to a deviation of the exact results with increasing $\theta$.

%%%%%%%%%%%%%%%%%%%%%%%%%%%%%%%%%%%%%
\subsection{Flow with topological freezing and complex frequencies}
\label{subsec:BreakdownLevelCrossing}

We have also investigated a variant of the flow with topological freezing: in this setup we exploit, that the topological term can be absorbed in an imaginary shift of the frequency. This potentially accommodates topological effects analogously to resurgence. For this purpose we rewrite the kinetic and topological terms in the initial action~\labelcref{eq:LagrangianUU} as
\begin{align}
   \int \frac{\df p}{2 \pi} \, \left[ \frac{m}{2} {\varphi}^* \, p^2 \, {\varphi} - \frac{i \theta}{2\pi} \varphi^* \, p \,{\varphi} \right]
    = 
        \int \frac{\df p}{2 \pi} \, \left[ \frac{m}{2} {\varphi}^* \, p_\theta^2  \, {\varphi} + \frac{\theta^2}{2m(2\pi)^2} \varphi^*{\varphi} \right]\,,
        \label{eq:initial action no.1}
\end{align}
with the complex frequency 
\begin{align}
p_\theta\equiv p + i\tilde{\theta} \equiv p + i\frac{\theta}{2m \pi}\,.
\label{eq:defOfptheta}
\end{align}
This is analogous to the system with a chemical potential if $\mu\equiv \tilde\theta$.
The potential receives an additional contribution proportional to $\theta^2$ from the second term in \labelcref{eq:initial action no.1} and then the initial potential is
given by
\begin{align}
    V_{k=\Lambda}(r^2) = \frac{m}{2} \tilde{\theta}^2 r^2 + \frac{g}{4} (r^2-1)^2\,.
    \label{eq:UV potential}
\end{align}

In such a setup, the system may be interpreted as a system with the polynomial potential \labelcref{eq:UV potential} at finite density. 
Here, we may make an ansatz for the effective action as
\begin{align}
    \Gamma_k 
    = \int \frac{\df p}{2\pi} \left[ \frac{m}{2} \varphi^* \,  p_\theta^2 \, \varphi  \right] + \int_\tau V_k(r^2)\,.
    \label{eq:GammakWithoutTopResum}
\end{align}
In this case, the frequency integral of the action has been extended into the complex plane due to the complex momentum \labelcref{eq:defOfptheta}. This is similar to theories with a chemical potential. There, the Silver Blaze property requires the use of regulators that do not destroy the underlying complex structure. 

We have refrained from such an investigation here as it goes beyond the scope of the present work. 
%
%%%%%%%%%%%%%%%%%%%%%%
\begin{figure}
    \centering
    \includegraphics[width=0.46\linewidth]{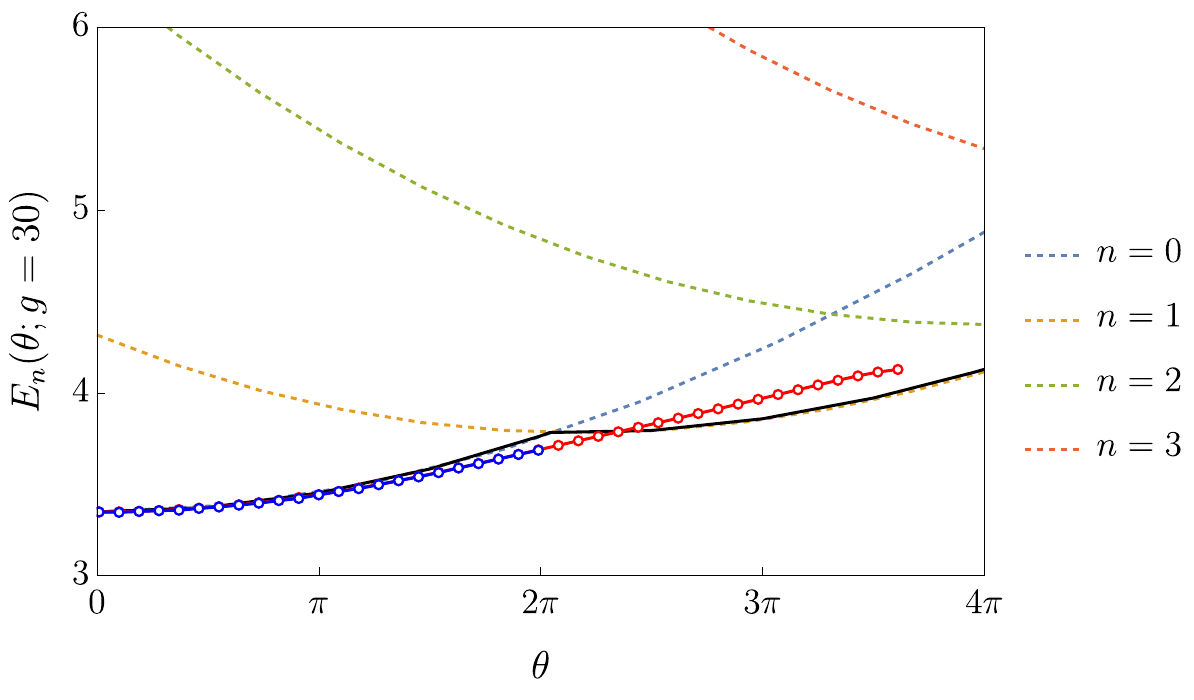}
    \hspace{4ex}
    \includegraphics[width=0.46\linewidth]{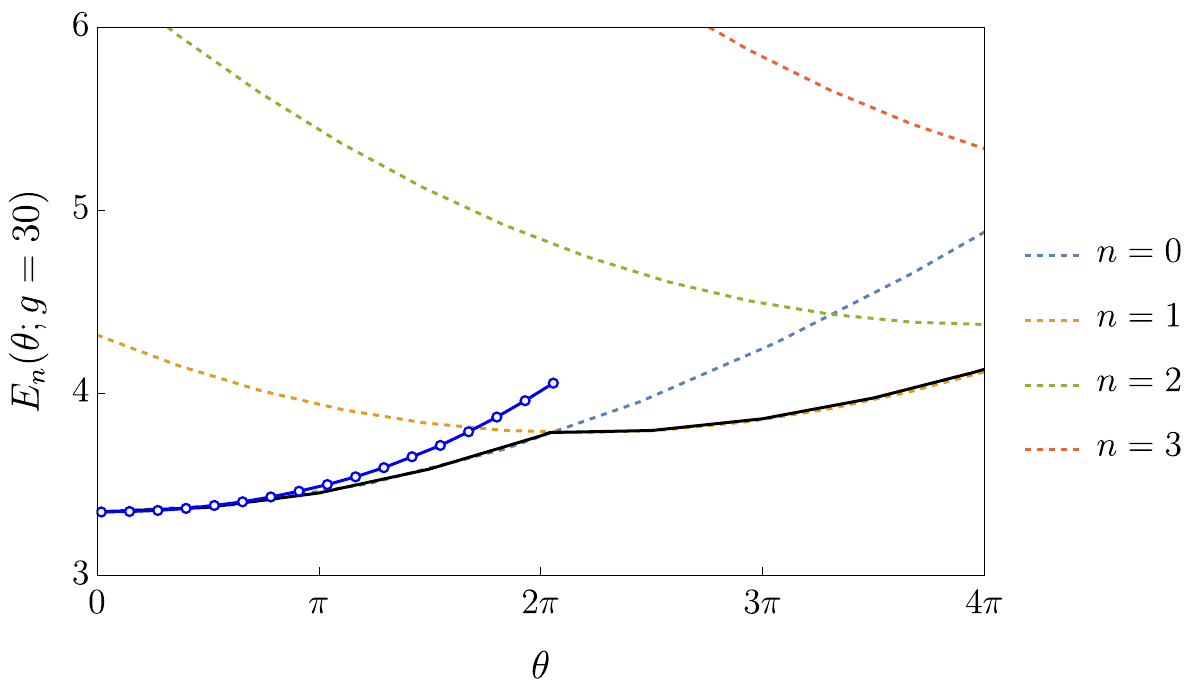}
    \caption{Energy eigenvalues for $n=0,\,\cdots,\,3$. The dashed-circle line exhibits the ground-state energy in terms of the $\theta$-parameter from \labelcref{eq:FlowVk+theta-term} (left) and \labelcref{eq:LitimComplexFlow} (right). 
    The dashed-colored lines show the solutions to the Schr\"odinger equation~\labelcref{eq:RadialSchroedinger}, and the solid-black line denotes the result obtained from the flow equation with the Litim-type regulator.
    }
    \label{fig:LitimRealFlowEnergy}
\end{figure}
%%%%%%%%%%%%%%%%%%%%%%
% 
Instead, we use a Litim-type regulator for the real part of the dispersion which is as close as possible to the regulator used in the main body of this work. We define 
\begin{align}
     R_k(p_\theta) = \operatorname{Re}(k^2-p_\theta^2) \Theta\bigl(\operatorname{Re}(k^2-p_\theta^2) \bigr)
     =\left(k^2 + \tilde{\theta}^2 - p^2\right) \, \tilde \Theta\left(k^2 + \tilde{\theta}^2 - p^2\right) 
     \,, 
     \label{eq:LitimReal}
\end{align}
where 
\begin{align} 
\tilde \Theta\left(k^2 + \tilde{\theta}^2 - p^2\right)=\Theta\bigl(\operatorname{Re}(k^2-p_\theta^2) \bigr)\,.
\label{eq:tildeTheta}
\end{align}
\Cref{eq:tildeTheta} implies 
\begin{align}
    \partial_t R_k(p_\theta) &= 2 k^2\, \tilde \Theta\left(k^2 + \tilde{\theta}^2 - p^2\right) \,,
    \label{LitimRegDerComplex}
\end{align}
and $p_\theta^2 + R_k(p_\theta)$ is given by 
\begin{align}\nonumber 
    p_\theta^2 + R_k(p_\theta)
     &= p^2 - \tilde{\theta}^2 + 2 i \tilde{\theta} p + (k^2 + \tilde{\theta}^2 - p^2) \tilde \Theta\left(k^2 + \tilde{\theta}^2 - p^2\right)\\[1ex]
    &= 2 i \tilde{\theta} p + k^2 \,\tilde \Theta\left(k^2 + \tilde{\theta}^2 - p^2\right)
   + \left( p^2 - \tilde{\theta}^2 \right)\, \tilde \Theta\left(p^2 - k^2 - \tilde{\theta}^2\right)\,.
\label{eq:LitimComplexKinetic}
\end{align}
The last term of the second line in \labelcref{eq:LitimComplexKinetic} does not contribute to the flow of the effective potential as it lies outside the range of the momentum integral with $p^2 \leq k^2+\tilde\theta^2$ following from  \labelcref{LitimRegDerComplex}. The flow equation is given by
\begin{align}
    \partial_t V_k = \int_\Omega \frac{\df p}{2 \pi} \frac{m k^2 \left[ \frac{m}{2} (k^2 + 2 i \tilde{\theta} p) + V^\prime + r^2 V^{\prime\prime} \right]}{\left[ \frac{m}{2} (k^2 + 2 i \tilde{\theta} p) + V^\prime + r^2 V^{\prime\prime} \right]^2 - \left[ r^2 V^{\prime\prime} \right]^2}\,,
\end{align}
in which the range of integration is $\Omega = \left[-\sqrt{k^2 + \tilde{\theta}^2},\sqrt{k^2 + \tilde{\theta}^2}\right]$.
Performing the momentum integral and taking the real part, we obtain
\begin{align}
    &\partial_t V_k(r^2)
    = \frac{k^2}{2 \pi |\tilde{\theta}|}\Bigg[  \arctan\left( \frac{2  |\tilde{\theta}|\sqrt{ k^2+\tilde{\theta}^2} }{c^{(1)}_k}\right)
    + \arctan\left(\frac{2 |\tilde{\theta}|\sqrt{k^2+\tilde{\theta}^2 } }{c^{(2)}_k }\right) \Bigg] \,,
\label{eq:LitimComplexFlow}
\end{align}    
where $c^{(1)}_k$ and $c^{(2)}_k$ are defined in \labelcref{eq:C1andC2}.

Now we solve the flow equation for $V_k(r^2)$ with the UV boundary condition \labelcref{eq:UV potential} at $k=\Lambda$.
In the right-hand side panel of \Cref{fig:LitimRealFlowEnergy}, we show the result of the vacuum energy in terms of the $\theta$-parameter by setting $m=1$. We have not taken $r^2$-derivatives and hence the numerical solution cannot be extended easily beyond $\omega=2 \pi$. As already discussed in \labelcref{sec:FlowEquationInRealVariableBasis}, this can be remedied by taking $r^2$-derivatives of \labelcref{eq:LitimComplexFlow}. This goes beyond the scope of the present work. 

The solution of the flow starts to deviate from the exact results with increasing $\theta$. However, the complex structure ('Silver Blaze') of the theory is violated by the regulator and this may be the (only) reason for the deviation. Accordingly, this setup deserves further analysis using regulators that accommodate the complex structure. A more comprehensive analysis will be presented elsewhere.

%%%%%%%%%%%%%%%%%%%%%%%%%%%%%%%%%%%%%%%
\section{RG evolution of the effective potential}
\label{app:RGevolutionofeffectivepotential}

\begin{figure}
    \centering
    \includegraphics[width=0.3\linewidth]{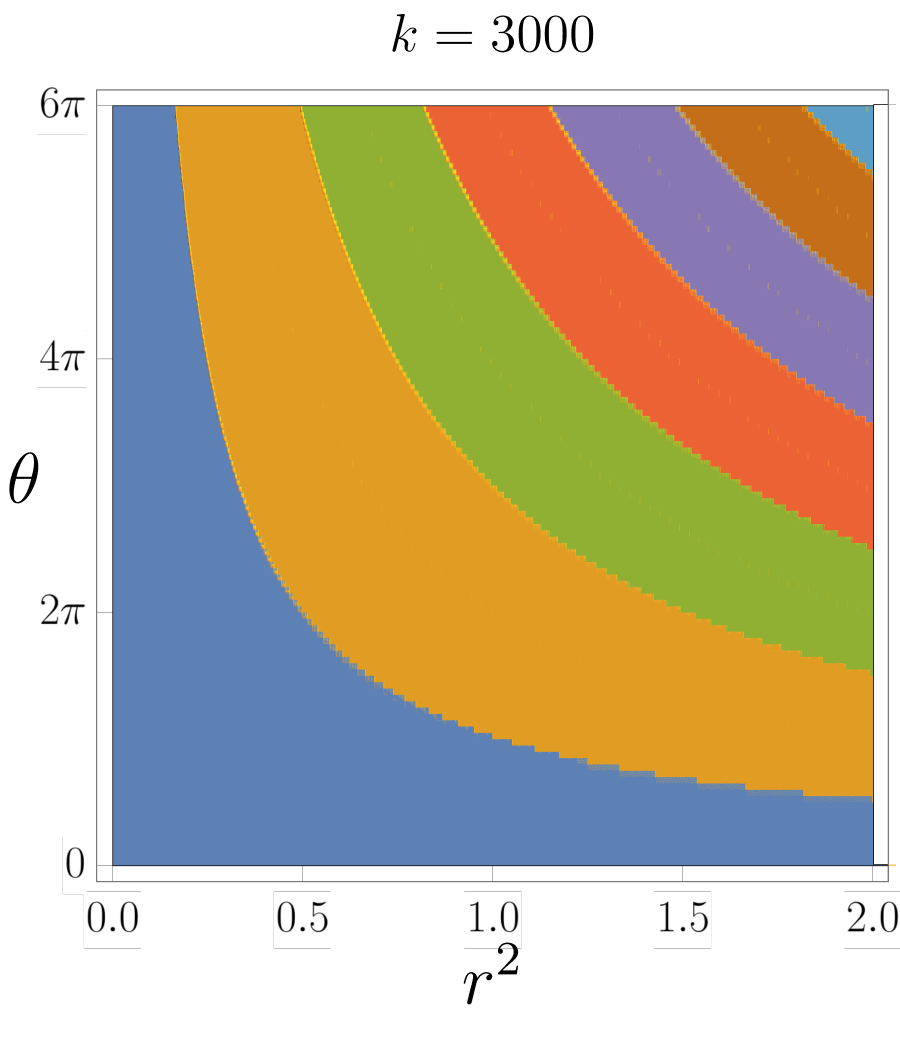}
    \hspace{5ex}
    \includegraphics[width=0.3\linewidth]{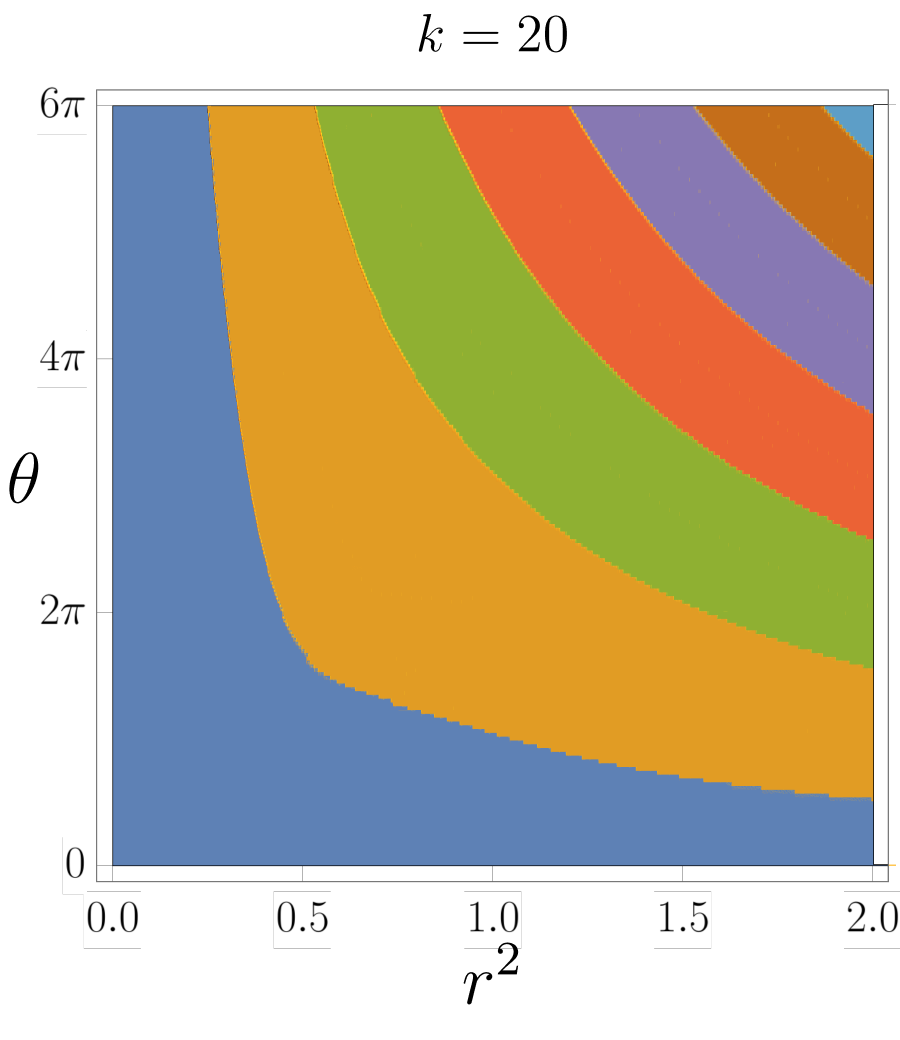}
    \hspace{5ex}
    \includegraphics[width=0.3\linewidth]{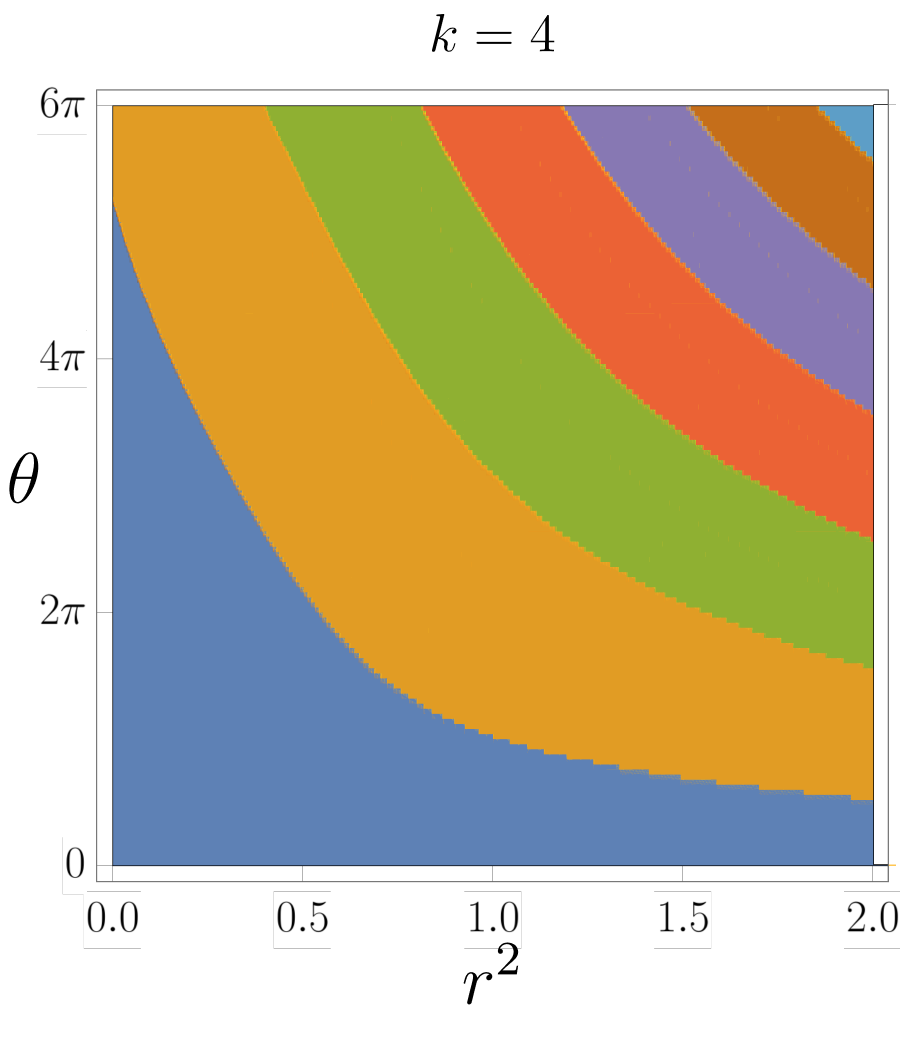}
    \hspace{5ex}
    \includegraphics[width=0.3\linewidth]{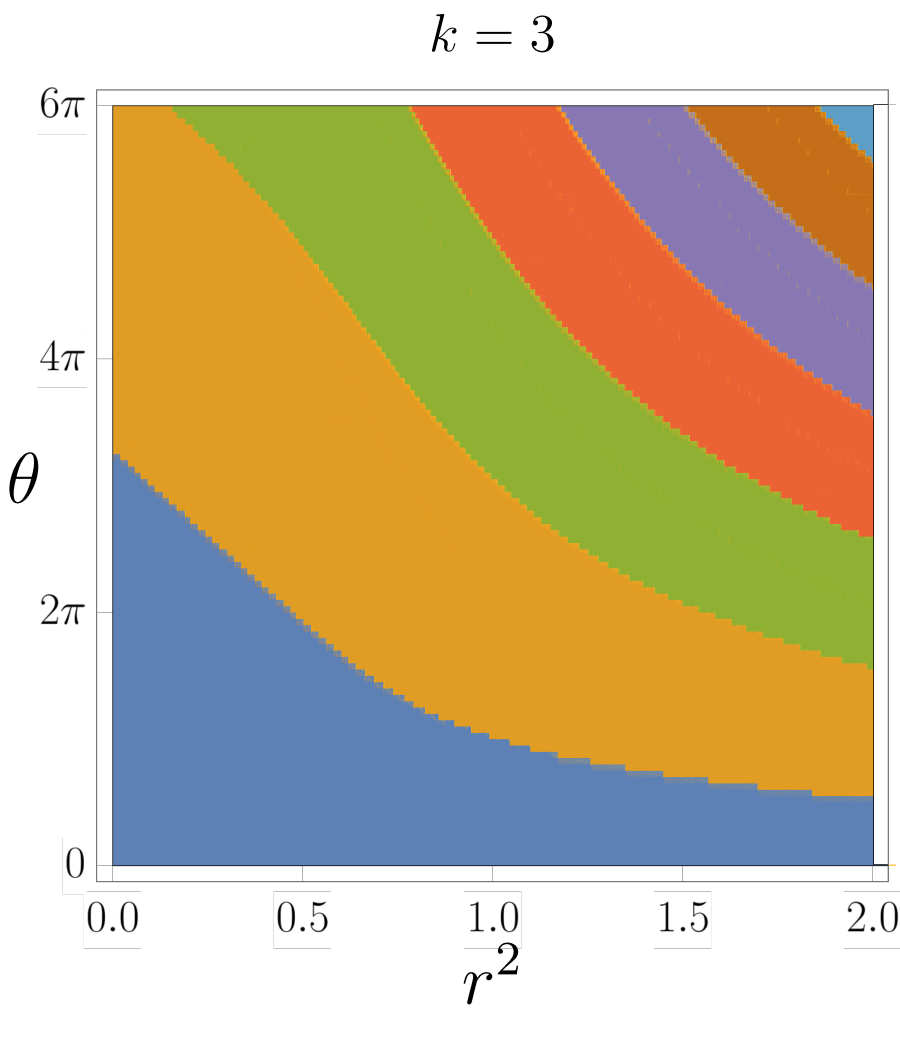}
      \hspace{5ex}
    \includegraphics[width=0.3\linewidth]{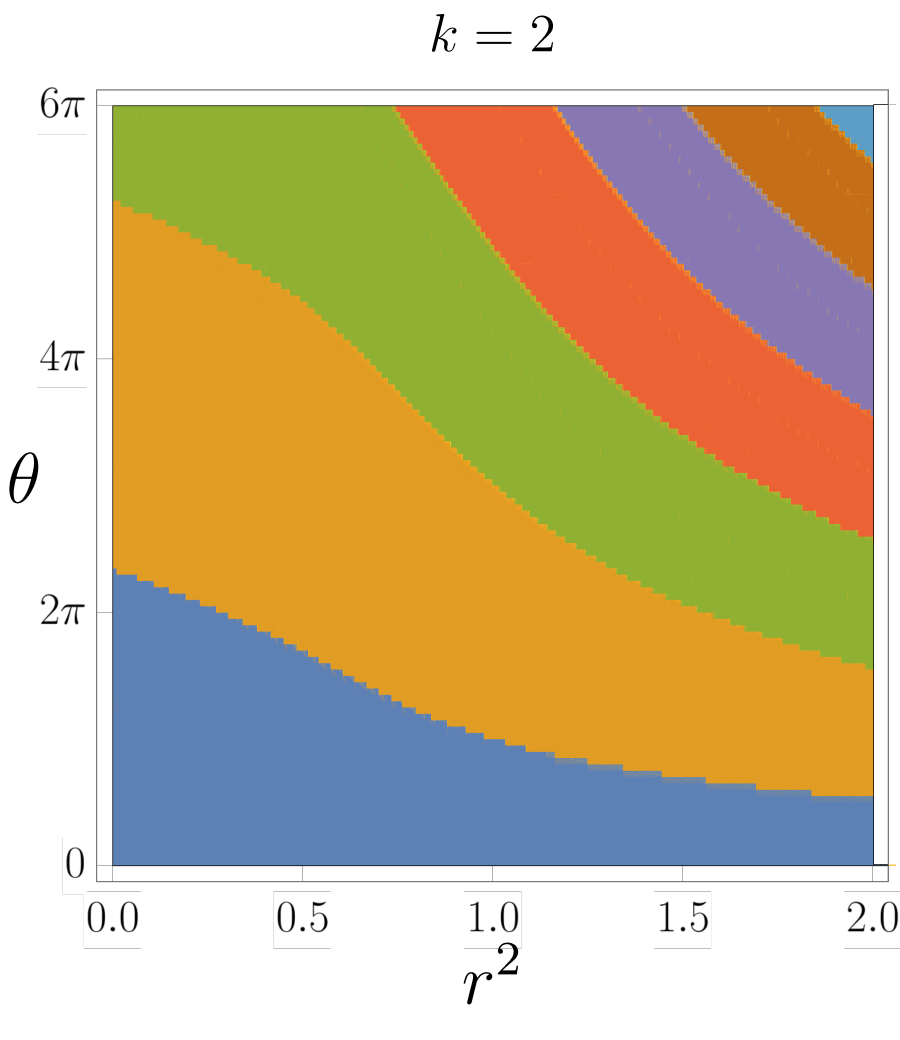}
      \hspace{5ex}
    \includegraphics[width=0.3\linewidth]{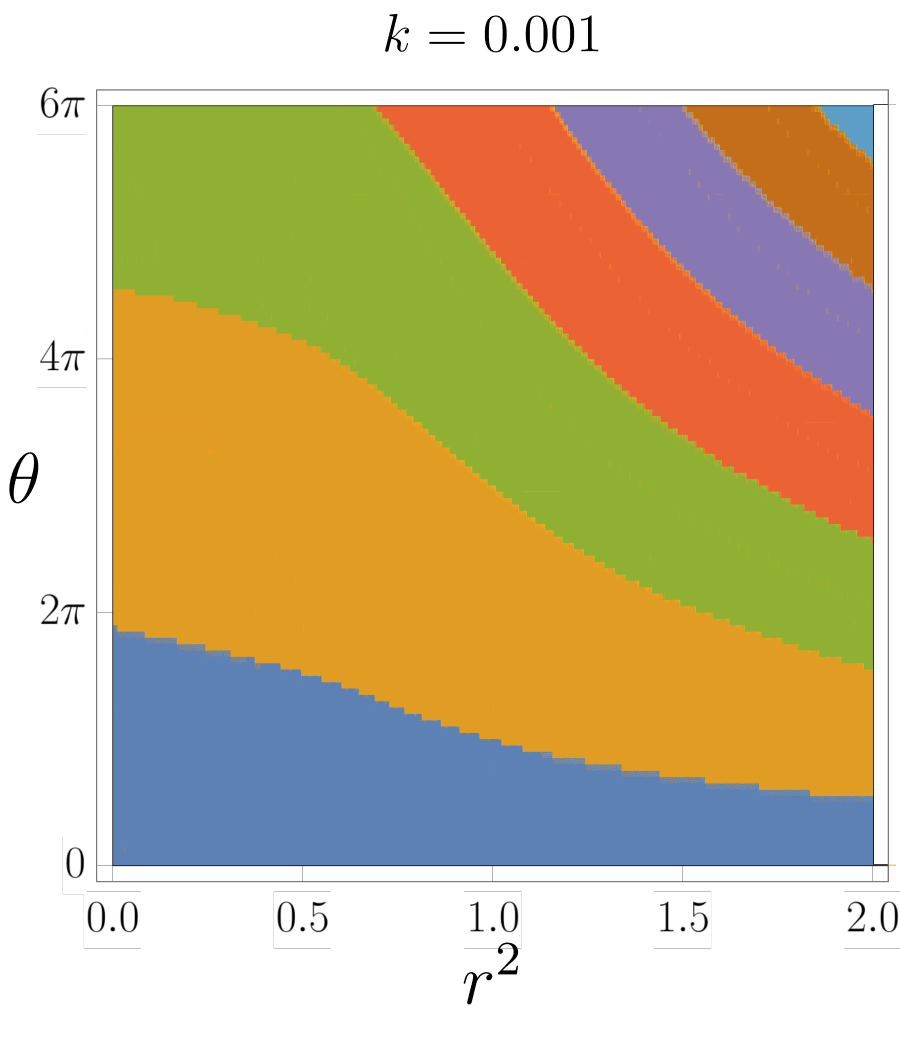}
    \caption{
    RG evolution of the effective potential on the $r^2$-$\theta$ plane.
    The RG scales are $k = 3000$ (top left), $k = 20$ (top right), $k = 4$ (middle left), $k = 3$ (middle right) and $k = 0.001$ (bottom).
    Different colors distinguish the energy levels which have minimal energy. From orange to blue regions, the energy level runs from $n=0$ to $n=6$. }
    \label{fig:jumpingFlowWithk}
\end{figure}

\begin{figure}
    \centering
    \includegraphics[width=0.4\linewidth]{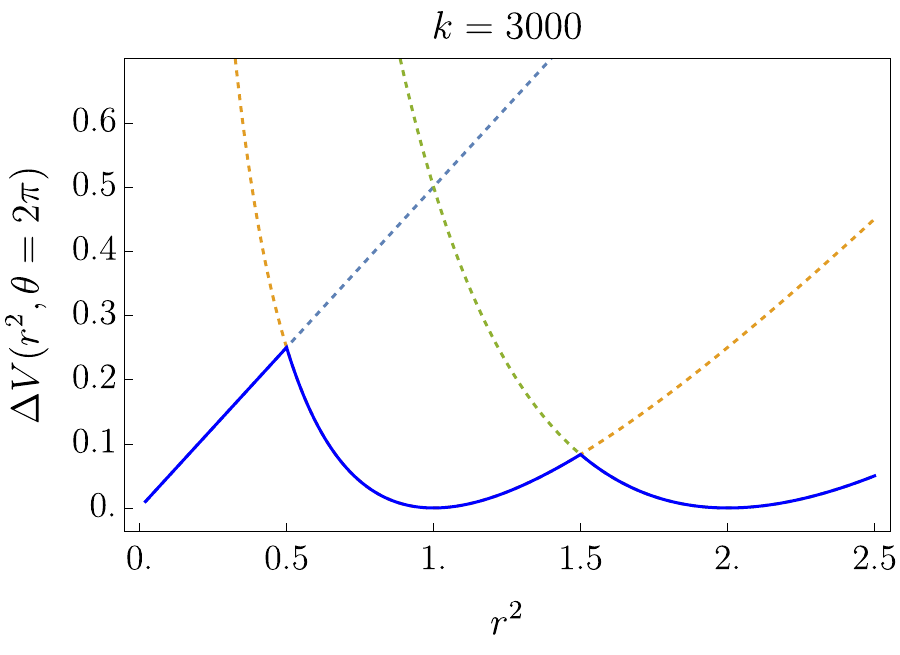}
    \hspace{5ex}
    \includegraphics[width=0.4\linewidth]{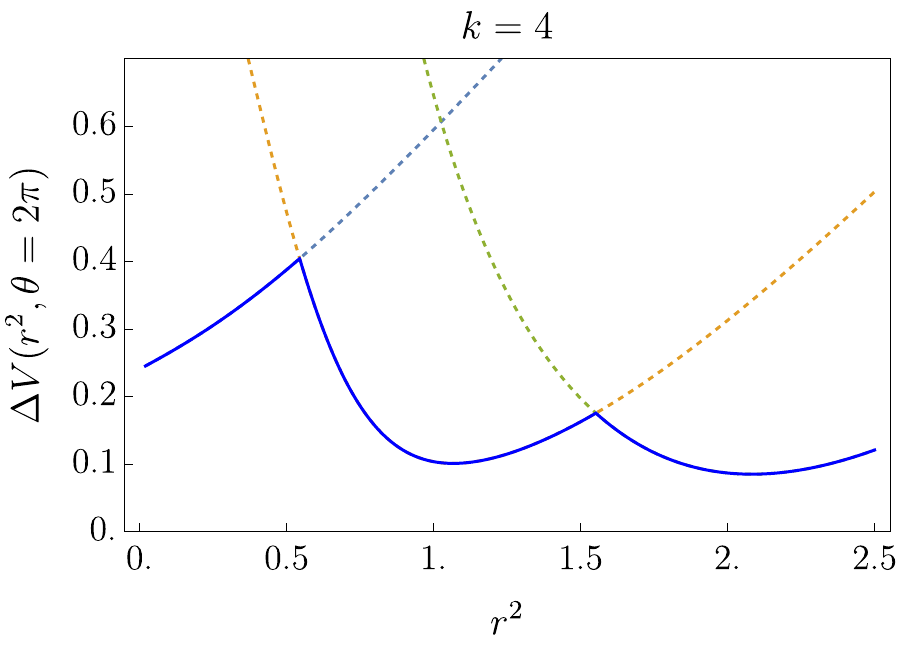}
    \hspace{5ex}
    \includegraphics[width=0.4\linewidth]{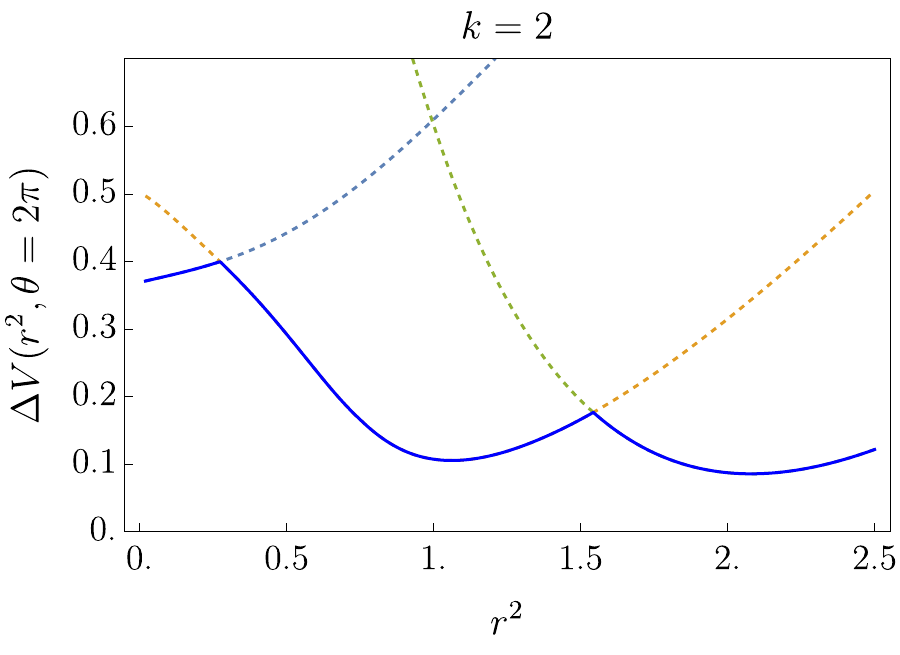}
    \hspace{5ex}
    \includegraphics[width=0.4\linewidth]{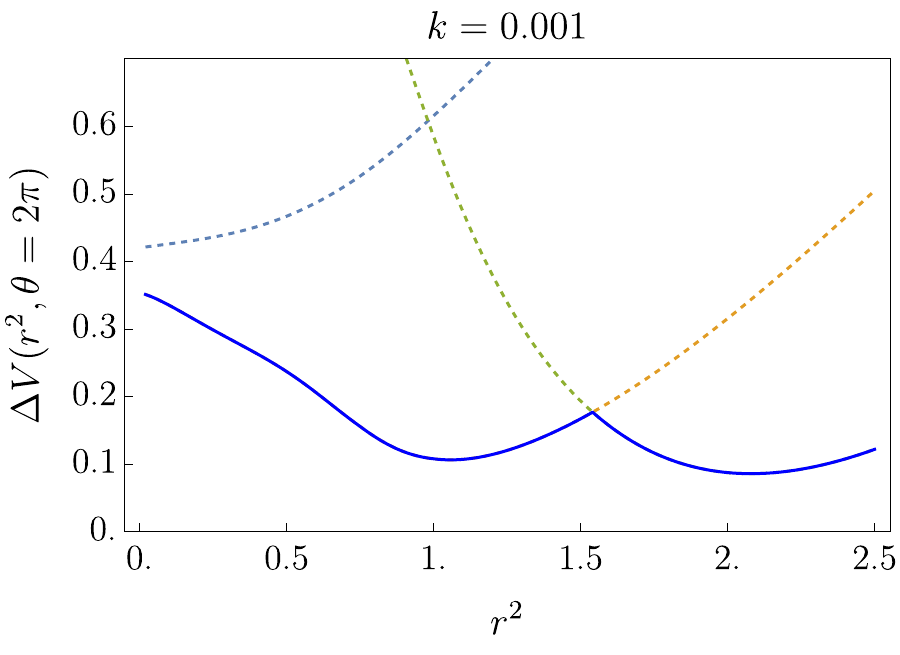}
    \caption{
    RG evolution of the effective potential with fixed $\theta = 2\pi$.
    The RG scales are $k = 3000$ (top), $k = 2$ (middle), $k = 0.001$ (bottom).
    Different colors of the dashed lines distinguish the energy levels, and the blue solid line is the total effective potential.}
    \label{fig:PotentialFixedTheta}
\end{figure}

In \Cref{fig:jumpingFlowWithk}, we show the RG-evolution of the effective potential for different RG scales $k$. The cusps are given by the boundary lines of the different colored areas. 
The effective potential at any scale $k$ automatically takes the minimum value among every energy levels $n$ in the $\beta\rightarrow \infty$ limit, and thus the level-crossing lines would move as long as the flow down to IR scale. In both cases, we see a good agreement with the fRG-results with the results from the Schr\"odinger equation at small-$\theta$ regime, with an increasing but still small deviation for $\theta \gsim 9 \pi$. This deviation originates in the simple LPA approximation. In \Cref{fig:PotentialFixedTheta} we show the effective potential energy evaluated at $\theta = 2\pi$ at different energy scales.

%

%%%%%%%%%%%%%%%%%%%%%%%%%%%%%%%%%%%
\section{Flow of the cusp position}
\label{app:CuspPostion}

In this Appendix we derive relations for the flowing cusp positions. The derivation is reminiscent of that of the flow of the flowing solution of the equations of motion. We keep this derivations minimal as we do not use these relations in the present work. Still we have checked their applicability numerically. 

The first relation works in the finite-volume case and its derivation resembles that of the flow of the cutoff dependent solution of the equation of motion: without loss of generality, we concentrate on one of the cusps with the flowing location $\varphi_k^\textrm{cusp}$. For the derivation of its flow we first consider the smooth case with a finite but large temporal volume, i.e., $\beta$ is finite but asymptotically large. In this case, the cusp is smoothed out. At the smoothed cusp locations, $\varphi_k^\textrm{cusp}$, the second derivative of the effective potential, $V_k^{(2)}$, takes a local maximum that diverges for $\beta \to \infty$. Hence, the third derivative vanishes at the location $\varphi^{\rm cusp}_k$, 
\begin{align} 
V_k^{(3)}(\varphi^{\rm cusp}_k)=0\,.
\label{eq:CuspLocation} 
\end{align}
Taking a total $t$-derivative leads us to 
\begin{align}
    \partial_t \varphi^{\rm cusp}_k = - \left. \frac{\partial_t V_k^{(3)}}{V_k^{(4)}} \right|_{\varphi^{\rm cusp}_k}\,.   \label{eq:cuspFlowFiniteVolume}
\end{align}
The numerator is simply the third derivative of the flow, evaluated at $\varphi_k^\textrm{cusp}$ with \labelcref{eq:CuspLocation}. All terms except one drop out because they are proportional to powers of $V_k^{(3)}(\varphi_k^\textrm{cusp})$. 

The first relation can be used directly in the infinite-volume case. Then, the cusp position is characterised by the crossing point of two different sectors, which reads
\begin{align}
    V_k^{-}(\varphi^{\rm cusp}_k) = V_k^{+}(\varphi^{\rm cusp}_k)\,,
    \label{eq:cuspCriterium1}
\end{align}
where the superscript $\pm$ denotes the potentials infinitesimally right or left of the cusp location. \Cref{eq:cuspCriterium1} holds true for all $k$ and hence its total $t$-derivative vanishes. This leads us readily to 
\begin{align}
    \partial_t \varphi^{\rm cusp}_k\Bigl[ V_k^{+,(1)}\left(\varphi^{\rm cusp}_k\right) - V_k^{-,(1)}\left(\varphi^{\rm cusp}_k\right) \Bigr] = - \Bigl[\partial_t V_k^{+} \left(\varphi^{\rm cusp}_k\right) - \partial_t V_k^{-}\left(\varphi^{\rm cusp}_k\right) \Bigr]\,.
    \label{eq:FlowCusp2pre}
\end{align}
We emphasise that \labelcref{eq:cuspCriterium1} holds true for any field value but for all but $\varphi^\textrm{cusp}$ both side in \labelcref{eq:FlowCusp2pre} vanish identically. In turn, for $\varphi_k^\textrm{cusp}$ they both are non-vanishing and we arrive at 
\begin{align}
    \partial_t \varphi^{\rm cusp}_k = - \left( \frac{\partial_t V_k^{+} - \partial_t V_k^{-}}{ V_k^{+,(1)} - V_k^{-,(1)}} \right)_{\varphi^{\rm cusp}_k}\,.
    \label{eq:FlowCusp2}
\end{align}
One may be tempted to use \labelcref{eq:FlowCusp2} in the finite-volume case where we have have $V^+_k=V^-_k$, but this leads us to \labelcref{eq:FlowCusp2pre} with vanishing left and right hand side, and \labelcref{eq:FlowCusp2} does not follow. Indeed, ignoring this fact leads us from \labelcref{eq:FlowCusp2} to 
\begin{align}
    \frac{\partial_t V_k^{+} - \partial_t V_k^{-}}{ V_k^{+, (1)} - V_k^{-, (1)}}  = \lim_{\epsilon \to 0} \frac{\partial_t V_k(\varphi + \epsilon) - \partial_t V_k(\varphi - \epsilon)}{ V_k^{(1)}(\varphi + \epsilon) - V_k^{(1)}(\varphi - \epsilon)} = \frac{\partial_t V_k^{(1)}}{V_k^{(2)}}\,.  
\label{eq:CuspFiniteVol}
\end{align}
\Cref{eq:CuspFiniteVol}  vanishes on the cusps in clear contradiction to the observed flow of $\varphi_k^\textrm{cusp}$ either given by  \labelcref{eq:cuspFlowFiniteVolume} or being extracted from the results. The vanishing of the right hand side of \labelcref{eq:CuspFiniteVol} originates in the fact that the flow of $V^{(1)}(\varphi_k^\textrm{cusp})$ is proportional to $V^{(3)}(\varphi_k^\textrm{cusp})=0$ and \labelcref{eq:CuspFiniteVol} vanishes. This concludes our discussion of the flow of $\varphi_k^\textrm{cusp}$.

\twocolumngrid

%%%%%%%%%%%%%%%%%%%%%%%%%%%%%%%%%%%%%
\bibliographystyle{JHEP} 
\bibliography{refs}

\end{document}